%% file: paper.tex
\documentclass[usenatbib]{mn2e}
\newcommand{\averng}{\bar{n}_{\rm g}}
\newcommand{\avernc}{\bar{n}_{\rm c}}
\newcommand{\averns}{\bar{n}_{\rm s}}
\newcommand{\avernx}{\bar{n}_{\rm x}}
\newcommand{\averrho}{\bar{\rho}_{\rm m}}

\newcommand{\rmd}{{\rm d}}
\newcommand{\rmg}{{\rm g}}
\newcommand{\rmm}{{\rm m}}
\newcommand{\rmh}{{\rm h}}
\newcommand{\rmc}{{\rm c}}
\newcommand{\rms}{{\rm s}}
\newcommand{\rmx}{{\rm x}}
\newcommand{\calI}{{\cal I}}
\newcommand{\calR}{{\cal R}}
\newcommand{\hatb}{\hat{b}}
\newcommand{\wigb}{\tilde{b}}
\newcommand{\sigb}{\sigma_{\rm b}}
\newcommand{\bvar}{b_{\rm var}}

\def\bxi{b_\xi}
\def\bvar{b_{\rm var}}

\def\b1{b_{1}}

\def\ltsima{$\; \buildrel < \over \sim \;$}
\def\lsim{\lower.5ex\hbox{\ltsima}}
\def\gtsima{$\; \buildrel > \over \sim \;$}
\def\gsim{\lower.5ex\hbox{\gtsima}}
\def\ga{\mathrel{\hbox{\rlap{\hbox{\lower4pt\hbox{$\sim$}}}\hbox{$>$}}}}

\input{psfig}
\input{macros}
\begin{document}
%

\title[Galaxy Biasing in the Halo Model]
      {On Combining Galaxy Clustering and Weak Lensing to Unveil 
       Galaxy Biasing via the Halo Model}

\author[Cacciato et al.]
       {\parbox[t]{\textwidth}{
        Marcello Cacciato$^{1}$\thanks{Minerva Fellow
        \newline 
        E-mail: cacciato@phys.huji.ac.il}, 
        Ofer Lahav$^{2}$,
        Frank C. van den Bosch$^{3}$,
        Henk Hoekstra$^{4}$,
        Avishai Dekel$^{1}$       
        } \\
           \vspace*{3pt} \\
         $^1$Racah Institute of Physics, The Hebrew University, Jerusalem 91904, Israel\\
        $^2$Depertment of Physics and Astronomy, University College London,            
            Gower Street, London, WC1EBT, UK\\
            $^3$Department of Astronomy, Yale University, PO. Box 208101, 
            New Haven, CT 06520-8101\\
            $^4$Leiden Observatory, Leiden University, Niels Bohrweg 2, 
            NL-2333 CA Leiden, The Netherlands
            }


\date{}

\pagerange{\pageref{firstpage}--\pageref{lastpage}}
\pubyear{2012}

\maketitle

\label{firstpage}


\begin{abstract} 
    We formulate the concept of non-linear and stochastic galaxy
    biasing in the framework of halo occupation statistics.  Using
    two-point statistics in projection, we define the galaxy bias
    function, $b_\rmg(r_\rmp)$, and the galaxy-dark matter
    cross-correlation function, ${\cal R}_{\rm gm}(r_\rmp)$, where
    $r_\rmp$ is the projected distance. We use the analytical halo
    model to predict how the scale dependence of $b_\rmg$ and
    $\calR_{\rm gm}$, over the range $0.1 h^{-1}\Mpc \lta r_\rmp \lta
    30 h^{-1}\Mpc$, depends on the non-linearity and stochasticity in
    halo occupation models.  In particular we quantify the effect
    due to the presence of central galaxies, the assumption for the
    radial distribution of satellite galaxies, the richness of the
    halo, and the Poisson character of the probability to have a
    certain number of satellite galaxies in a halo of a certain
    mass. Overall, brighter galaxies reveal a stronger scale
    dependence, and out to a larger radius. In real-space, we find
    that galaxy bias becomes scale independent, with $\calR_\rmg = 1$,
    for radii $r \geq 1 - 5 h^{-1} \Mpc$, depending on luminosity.
    However, galaxy bias is scale-dependent out to much larger radii
    when one uses the projected quantities defined in this paper.
    These projected bias functions have the advantage that they are 
    more easily accessible observationally and that their scale dependence
    carries a wealth of information regarding the properties
    of galaxy biasing.
    To observationally constrain the parameters of the halo occupation statistics and 
    to unveil the origin of galaxy biasing we propose the use of the bias function
    $\Gamma_{\rm gm}(r_{\rm p}) \equiv b_{\rm g}(r_{\rm p})/{\cal R}_{\rm gm}(r_{\rm p})$.
    This function is obtained via a combination of weak gravitational lensing and galaxy clustering, 
    and it can be measured using existing and forthcoming 
    imaging and spectroscopic galaxy surveys.
\end{abstract}




\section{Introduction}
\label{sec:intro}

According to the current paradigm of structure formation in the
Universe, galaxies form and reside within dark matter haloes which
emerge from the (non-linear) growth of primordial density
fluctuations.  In this scenario, one expects that the spatial
distribution of galaxies traces, to first order, that of the
underlying dark matter. However, due to the complexity of galaxy
formation and evolution, galaxies are also expected to be somewhat
biased tracers. In general, the relationship between the galaxy
and dark matter distribution is loosely referred to as {\it
  galaxy biasing} (see e.g. \citealt{1985ApJ...292..371D,
  1986ApJ...304...15B,1987Natur.326..455D}). At any cosmic epoch, this
relation is the end product of processes such as non-linear
gravitational collapse, gas cooling, star formation, and possible
feedback mechanisms.  Therefore, understanding the features of galaxy
biasing is crucial for a thorough comprehension of galaxy formation
and evolution as well as for the interpretation of those studies which
use galaxies as tracers of the underlying dark matter distribution in
an attempt to constrain cosmological parameters.

The simplest way to model galaxy biasing is by assuming a linear
and deterministic relation between the matter and the galaxy density
fields.  Galaxy formation, though, is a complicated process and the
validity of such a simplistic assumption is highly questionable. As
a consequence, numerous authors have presented various arguments for
considering modifications from a simple linear and deterministic
biasing scheme (e.g., \citealt{1984ApJ...284L...9K,
1985ApJ...292..371D, 1986ApJ...304...15B, 1986ApJ...303...39D,
1987Natur.326..455D, 1988ApJ...328...34B, 1991MNRAS.253P..31B,
1992ApJ...396..430L, 1998ApJ...504..607S, 1998ApJ...500L..79T,
1999ApJ...520...24D, 1999ApJ...522...46T}). In addition,
cosmological simulations and semi-analytical models of galaxy
formation strongly suggest that galaxy bias indeed takes on a
non-trivial form (e.g., \citealt{1992ApJ...399L.113C,
1997MNRAS.286..795K, 1999ApJ...518L..69T, 2001MNRAS.320..289S}).
From the observational side, numerous attempts have been made to
test whether galaxy bias is linear and deterministic. This includes
studies that compare the clustering properties of different samples
of galaxies (e.g., \citealt{1999ApJ...518L..69T,
2000ApJ...544...63B, 2005MNRAS.356..456C, 2005MNRAS.356..247W,
2007ApJ...664..608W, 2008MNRAS.385.1635S, 2011ApJ...736...59Z}),
studies that measure higher-order correlation functions (e.g.,
\citealt{1999ApJ...521L..83F, 2002ApJ...570...75S,
2002MNRAS.335..432V}), studies that compare observed fluctuations
in the galaxy distribution to matter fluctuations predicted in
numerical simulations (e.g., \citealt{2005A&A...442..801M,
2011ApJ...731..102K}), and studies that combine galaxy clustering
measurements with gravitational lensing measurements
(\citealt{2001ApJ...558L..11H, 2002ApJ...577..604H,
2003MNRAS.346..994P, 2004AJ....127.2544S, 2007A&A...461..861S,
2012arXiv1202.6491J}). We refer the reader to
\S\ref{sec:observations} for a detailed discussion of the pros and
cons of these different methods. Here we simply mention that the
majority of these observational studies have confirmed that galaxy
bias is neither linear nor deterministic.

Although this observational result is in qualitative agreement
with theoretical predictions, we still lack a {\it direct} link
between model predictions and actual measurements. This is mainly a
consequence of the fact that the standard formalism used to define
(the non-linearity and stochasticity of) galaxy bias is difficult to
interpret in the framework of galaxy formation models. In this paper
we introduce a new methodology that allows for a more intuitive
interpretation of galaxy bias that is more directly linked to
various concepts of galaxy formation theory. In particular, we
reformulate the parameterization introduced by Dekel \& Lahav (1999;
hereafter DL99) to describe the non-linearity and stochasticity of
the relation between galaxies and matter in the langauge of halo
occupation statistics. Since galaxies are believed to reside in dark
matter haloes, halo occupation distributions are the natural way to
parameterize the galaxy-dark matter connection, and thus the concept
of galaxy bias. In addition, combined with the halo model, which is
an analytical formalism to describe the non-linear clustering of dark matter 
(e.g., \citealt{2000ApJ...543..503M,2000MNRAS.318..203S}), 
halo occupation models can be used to
compute the $n$-point statistics of the galaxy distribution (see
e.g., \citealt{2002PhR...372....1C}), which can be compared with
observations.  Hence, while the formalism in DL99 was not designed
to identify the sources of stochasticity and non-linearity in the
relation between galaxies and dark matter, its reformulation in
terms of halo occupation statistics holds the potential to unveil
{\it the hidden factors} from which deviations from the simple
linear and deterministic galaxy biasing arise\footnote{In DL99,
sources of deviations from the linear and deterministic biasing
scheme are repeatedly referred to as {\it ``hidden factors
affecting galaxy formation''}.}. In order to demonstrate the
potential power of this methodology, we make extensive use of
two-point statistics (galaxy-galaxy, galaxy matter, and
matter-matter correlation functions) to investigate how
non-linearity and stochasticity in the halo occupation statistics,
which have intuitive connections with galaxy formation, impact 
the observable properties of galaxy bias.

The study presented in this paper exploits the fact that with the
advent of large and homogeneous galaxy surveys, it has become possible
to conduct statistical studies of galaxy properties in connection with
the assumed underlying dark matter distribution.  For instance, one
has accurate measurements of the two-point correlation function of
galaxies as a function of their properties, such as luminosity,
morphology and colour (e.g.  \citealt{2000A&A...355....1G, N01,N02,
  2005ApJ...630....1Z,2007ApJ...664..608W}) which are routinely
compared with the clustering properties of simulated dark matter
haloes of different masses.  Alongside with clustering measurements,
gravitational lensing represents a direct diagnostic of the
galaxy-dark matter connection.  For instance, galaxy-galaxy lensing
and cosmic shear (which are related to the galaxy-matter and the
matter-matter correlation functions) have rapidly evolved into
techniques capable of probing the properties of the dark matter
distribution using galaxies as tracers.  There is growing scientific
interest in combining the complementary information obtained from weak
lensing and galaxy clustering to further constrain the properties of
the galaxy-dark matter connection and in turn the underlying
cosmological model.  For instance, \citet{2011arXiv1109.4852G}
considered three different types of probes in their analysis: 1)
angular clustering from galaxy-galaxy autocorrelation in narrow
redshift bins; 2) weak lensing from shear-shear, galaxy-shear and
magnification (i.e. galaxy-matter cross-correlation); and 3) redshift
space distortions from the ratio of transverse to radial modes. The
combination of such measurements provides a significant improvement in
the forecast for the evolution of the dark energy equation of state
and the cosmic growth evolution.  This improvement comes from
measurements of galaxy bias, which affects both redshift-space
distortion and weak lensing cross-correlations, but in different ways
(see also \citealt{2006ApJ...652...26Y,2009MNRAS.394..929C}).  A
similar approach can also be used to test the validity of general
  relativity (GR) on cosmological scales, provided that the
scale-dependence of galaxy bias due to stochasticity and non-linearity
is sufficiently small (e.g.,
\citealt{2007PhRvL..99n1302Z,2010Natur.464..256R}).

In light of forthcoming surveys such as Pan-STARRS, KIDS, DES, LSST,
and Euclid\footnote{Pan-STARRS:Panoramic Survey Telescope \& Rapid
  Response System, http://pan-starrs.ifa.hawaii.edu; \\ KIDS:
  KIlo-Degree Survey, \\ http://www.astro-wise.org/projects/KIDS/; \\
  DES: Dark Energy Survey, https://www.darkenergysurvey.org;\\ LSST:
  Large Synoptic Survey Telescope, http://www.lsst.org; \\ Euclid:
  http://www.euclid-ec.org} (\citealt{2011arXiv1110.3193L}) 
which will produce deep imaging over
large fractions of the sky and thereby yield statistical measurements
with unprecedented quality, it is mandatory to improve our
understanding of galaxy bias which currently limits our ability to
exploit the potential of such surveys. To this aim, the approach
undertaken in this paper consists of introducing a method which
predicts features due to galaxy biasing observable via a combination
of galaxy clustering and galaxy lensing measurements.  One spin-off of
this paper is a characterisation of the length scale above which it is
safe to assume that galaxy bias is scale-independent. This is
important for a number of cosmological studies, such as testing 
  GR via the method advocated by \citet{2007PhRvL..99n1302Z}
  and used by \citet{2010Natur.464..256R}.

This paper is organized as follows.  In \S2 we re-visit the classical
concepts of galaxy biasing and we also formulate it in the context of
the halo model.  In \S3 we present a realistic description of the halo
occupation statistics and we describe how its assumptions affect the
value of galaxy bias parameters.  In \S4 we introduce the galaxy bias
functions $b_{\rm g}$, ${\cal R}_{\rm gm}$ and ${\Gamma_{\rm gm}}$ in
terms of two-point statistics. In \S5 we analyze different scenarios
of the way galaxies may populate dark matter haloes highlighting how
these scenarios translate in distinct features in the scale dependence
of the galaxy bias functions.  In \S6 we comment on existing
observational attempts to constrain galaxy bias.  In \S7 we discuss
our results and draw conclusions. Throughout this paper, we
assume a flat $\Lambda$CDM cosmology specified by the following
cosmological parameters: $(\Omega_{\rm m}, \sigma_8, n, h)=(0.24,
0.74, 0.95, 0.73)$ supported by the results of the third year of the
Wilkinson Microwave Anisotropy Probe (WMAP,
\citealt{2007ApJS..170..377S}).


\section{Galaxy Biasing}
\label{sec:galbias}

\citet{1999ApJ...520...24D} introduced a general formalism to describe
galaxy biasing. In particular, they introduced convenient parameters
to describe the non-linearity and stochasticity of the relation
between galaxies and matter. A downside of their formalism, however,
is that it is based on the {\it smoothed} galaxy and matter density
fields. This smoothing blurs the interpretation of the associated
two-point statistics on small scales (smaller than the smoothing
length). In addition, the various bias parameters introduced in DL99
do not have counterparts that are easily accessible from observations,
making it difficult to constrain the amounts of non-linearity and/or
stochasticty as defined by DL99.  In recent years, a more convenient
method for describing the bias and clustering properties of galaxies
has emerged in the form of halo occupation statistics (e.g.,
\citealt{1998ApJ...494....1J}; Seljak 2000;
\citealt{2000MNRAS.318.1144P,2002ApJ...575..587B,
  YMB03,2005MNRAS.361..415C,vdB07}). In this section, after a short
recap of the classical description of galaxy biasing, we reformulate
the DL99 formalism in the terminology of halo occupation statistics.
This formulation allows for a far more direct and intuitive
interpretation of the concepts of non-linearity and stochasticity.

\subsection{Classical Description}  
\label{sec:classical}

Let $n_{\rm g}({\bf x})$ and $\rho_{\rm m}({\bf x})$ indicate the {\it
  local} density fields of galaxies and matter at location ${\bf x}$,
respectively. The corresponding overdensity fields are defined as
\begin{equation}\label{eq:deltas}
\delta_{\rm g}({\bf x}) = 
\frac{n_{\rm g}({\bf x}) -{\bar n}_{\rm g}}{{\bar n}_{\rm g}} 
\qquad {\rm and} \qquad \delta_{\rm m}({\bf x}) = 
\frac{\rho_{\rm m}({\bf x}) -{\bar \rho_{\rm m}}}{{\bar \rho_{\rm m}}} \, ,
\end{equation}
where ${\bar n}_{\rm g}$ is the average number density of galaxies and
${\bar \rho_{\rm m}}$ is the dark matter background density.  Both
these fields are smoothed with a smoothing window which defines the
term {\it `local'}. Galaxy bias is said to be {\it linear and
  deterministic} if
\begin{equation}\label{eq:ld}
\delta_{\rm g}({\bf x}) = b_{\rm g} \delta_{\rm m}({\bf x}) \, ,
\end{equation}
where $b_{\rm g}$ is referred to as the galaxy bias parameter.
Clearly, such a biasing scheme is highly idealized. As emphasized in
DL99, the assumption of linear deterministic biasing must brake down
in deep voids if $b_\rmg > 1$ simply because $\delta_\rmg \geq
-1$. Furthermore, numerical simulations have shown that the bias of
dark matter haloes is both non-linear and stochastic (e.g.,
\citealt{1996MNRAS.282..347M,2001MNRAS.320..289S}). Hence, it is only
natural that galaxies, which reside in dark matter haloes, also are
biased in a non-linear and stochastic manner.  Indeed, simulations of
galaxy formation in a cosmological context suggest a biasing relation
that is non-linear, scale-dependent and stochastic (see
e.g. \citealt{2001MNRAS.320..289S}). Here scale-dependence refers to
the fact that the biasing description depends on the smoothing scale
used to define $\delta_\rmg$ and $\delta_\rmm$.

Based on these considerations, DL99 generalized the concept of galaxy
bias by considering the local biasing relation between galaxies and
matter to be a {\it random} process, specified by the biasing
conditional distribution, $P(\delta_\rmg|\delta_\rmm)$, of having a
galaxy overdensity $\delta_\rmg$ at a given $\delta_\rmm$.  They
defined the {\it mean biasing function}, $b(\delta_\rmm)$, by the
conditional mean:
\begin{equation}\label{mbf}
b(\delta_\rmm) \delta_\rmm  \equiv \langle \delta_\rmg|\delta_\rmm \rangle = 
\int \delta_\rmg \, P(\delta_\rmg|\delta_\rmm) \, \rmd\delta_\rmg
\end{equation}
The function $b(\delta_\rmm)$ allows for any possible non-linear
biasing and fully characterizes it, reducing to the special case of
linear biasing when $b(\delta_\rmm) = b_\rmg$ is a constant
independent of $\delta_\rmm$. The stochasticity of galaxy bias is
captured by the {\it random biasing field}, $\varepsilon$, which is
defined by
\begin{equation}\label{rbf}
\varepsilon \equiv \delta_\rmg - \langle \delta_\rmg|\delta_\rmm \rangle
\end{equation}
and has a vanishing local conditional mean, i.e., $\langle \varepsilon
| \delta_\rmm \rangle = 0$. The variance of $\varepsilon$ at a
given $\delta_\rmm$ defines the stochasticity function
$\langle\varepsilon^2|\delta_\rmm\rangle$, whose average over
$\delta_\rmm$ specifies the overall (global) stochasticity of the
galaxy field,
\begin{equation}\label{globalstoch}
\langle \varepsilon^2 \rangle = \int \langle \varepsilon^2|\delta_\rmm \rangle
\, P(\delta_\rmm) \, \rmd\delta_\rmm\,,
\end{equation}
with $P(\delta_\rmm)$ the probability distribution function (PDF) of
$\delta_\rmm$. As mentioned in DL99, the quantity $\varepsilon$ serves
as an analytical tool to account for stochasticity without identifying
its sources. The formalism presented in the next section gives a
natural framework within which the {\it hidden} sources of
stochasticity can be unveiled. In general, stochasticity is expected
to arise from: i) the relation between dark matter haloes and the
underlying dark matter density field; and ii) the way galaxies
populate dark matter haloes. The former is better addressed using
  cosmological N-body simulations (see e.g.,
  \citealt{1996MNRAS.282..347M, 1998MNRAS.297..692C,
    1999ApJ...513L..99P, 1999MNRAS.304..767S}) and is not the goal
of this paper. Rather, we focus on the second source of stochasticity,
which we address using the analytical halo model complemented with
halo occupation statistics.

\subsection{Non-linearity and Stochasticity of Halo Occupation Statistics}
\label{sec:newform}

We now reformulate the DL99 formalism in the language of halo
occupation statistics. Rather than using the overdensities
$\delta_\rmg$ and $\delta_\rmm$ of the smoothed galaxy and matter
density fields, we use two new variables: the number of galaxies in a
dark matter halo, $N$, and the mass of that dark matter halo, $M$. In
particular, the relation between galaxies and dark matter is now
described by the halo occupation distribution $P(N|M)$, rather than
the conditional distribution $P(\delta_\rmg|\delta_\rmm)$.  The
equivalent of the mean biasing function, $b(\delta_\rmm)$, defined in
Eq.~(\ref{mbf}), now becomes
\begin{equation}\label{bMdef}
b(M) \equiv {\averrho \over \averng} \, {\langle N|M \rangle \over M} \,, 
\end{equation}
where $\langle N |M \rangle$ is the mean of the halo occupation
distribution, i.e.,
\begin{equation}\label{NMaver}
\langle N|M\rangle = \sum_{N=0}^{\infty} N \, P(N|M)\,,
\end{equation}
and the factor $\averrho/\averng$ is required on dimensional grounds.
Following the same nomenclature as in DL99, {\it linear, deterministic
  biasing} now corresponds to
\begin{equation}\label{lindetbias}
N = {\averng \over \averrho} \, M\,,
\end{equation}
which yields $b(M) = 1$. Since $N$ is an integer, whereas the
quantities on the rhs of Eq.~(\ref{lindetbias}) are real, it is
immediately clear that in our new formulation linear, deterministic
biasing is unphysical. Note, though, that this does not imply that
$b(M) =1$ is unphysical; after all, $b(M) = 1$ can be established by
having $\langle N |M\rangle = (\averng/\averrho) \, M$, which is
possible (in practice). In this case, however, there must be non-zero
stochasticity. If, on the other hand, there is a deterministic
relation between $N$ and $M$, then the bias cannot be linear (i.e.,
$b(M) \ne 1$).

Following DL99, we characterize the function $b(M)$ by the moments
$\hatb$ and $\wigb$ defined by
\begin{equation}\label{moments}
\hatb \equiv {\langle b(M) M^2 \rangle \over \sigma_M^2}\,, 
\qquad {\rm and} \qquad 
\wigb^2 \equiv {\langle b^2(M) M^2 \rangle \over \sigma_M^2}\,.
\end{equation}
Here $\langle ... \rangle$ indicates an average over dark matter
haloes, i.e.,
\begin{equation}\label{haloaver}
\langle A \rangle \equiv {\int A \, n(M) \, \rmd M \over \int n(M) \, \rmd M}
\end{equation}
with $n(M)$ the halo mass function, and $\sigma_M^2 \equiv \langle M^2
\rangle$. Galaxy bias is linear if $\wigb/\hatb = 1$. It is
straightforward to see that this is only possible if $b(M)$ is
independent of halo mass. Hence, in our new formulation we have that
linear bias corresponds to halo occupation statistics for which
$\langle N|M \rangle \propto M$.

Motivated by DL99, we define the {\it random halo bias}
\begin{equation}\label{epsNdef}
\varepsilon_N \equiv N - \langle N|M \rangle\,,
\end{equation}
for which the conditional mean vanishes, i.e., $\langle \varepsilon_N
|M\rangle = 0$.  The variance of $\varepsilon_N$ for halos of a given
mass defines the {\it halo stochasticity function}
\begin{equation}\label{bsfmod}
\sigb^2(M) \equiv \left({\averrho \over \averng} \right)^2 \, 
{\langle \varepsilon^2_N|M \rangle \over \sigma_M^2}
\end{equation}
where, following DL99, the scaling by $\sigma_M^2$ is introduced for
convenience.  By averaging over halos of all masses, we finally
obtain the {\it stochasticity parameter}
\begin{equation}\label{stochpardef}
\sigb^2 \equiv \left({\averrho \over \averng} \right)^2 \, 
{\langle \varepsilon^2_N \rangle \over \sigma_M^2}
\end{equation}
Galaxy bias is said to be deterministic if $\sigb = 0$.

In addition to the bias parameters $\hatb$ and $\wigb$, which are mass
moments of the mean biasing function $b(M)$, one can also define other
bias parameters.  In particular, DL99 introduced the ratio of the
variances, $\bvar \equiv \langle \delta^2_\rmg \rangle / \langle
\delta^2_\rmm \rangle$, which in our reformulation becomes
\begin{equation}\label{bvardef}
\bvar \equiv \left({\averrho\over\averng}\right)^2 \, 
{\sigma_N^2 \over \sigma_M^2} = \left({\averrho\over\averng}\right)^2 \,
{\langle N^2 \rangle \over \langle M^2 \rangle}\,.
\end{equation}
Using Eq.~(\ref{epsNdef}) and the fact that $\langle \varepsilon_N
\rangle = 0$, one finds that
\begin{equation}\label{bvaralt}
\bvar^2 = \wigb^2 + \sigb^2\,.
\end{equation}
This equation, which is exactly the same as in DL99, makes it explicit
that the bias parameter $\bvar$ is sensitive to both non-linearity and
stochasticity. Combining Eqs.~(\ref{bvardef}) and~(\ref{bvaralt}) we
have that
\begin{equation}\label{Nsquared}
\langle N^2 \rangle = \left({\averrho\over\averng}\right)^2 \,
\left[\wigb^2 + \sigb^2 \right] \, \sigma_M^2\,.
\end{equation}
It is useful to compare this to the {\it covariance}
\begin{equation}\label{NM}
\langle N M \rangle = {\averrho\over\averng} \, \hatb \, \sigma_M^2\,,
\end{equation}
which follows directly from Eqs.~(\ref{bMdef}) and~(\ref{moments}).
Unlike the variance $\langle N^2 \rangle$, the covariance has no
additional contribution from the biasing scatter $\sigb$ (see
also DL99).

Finally, we define the {\it linear correlation coefficient}
\begin{equation}\label{rdef}
r \equiv {\langle N M \rangle \over \sigma_N \, \sigma_M}\,.
\end{equation}
Using Eqs.~(\ref{Nsquared})-(\ref{NM}), it is straightforward to
see that we can write 
\begin{equation}\label{hatbint}
\hatb = \bvar \, r\,.
\end{equation}
Hence, the first moment of the mean bias function $b(M)$ is simply the
product of the ratio of variances, $\bvar$, and the linear correlation
coefficient, $r$.
 
Using these parameters, we can now characterize a few special cases.
As already mentioned above, the discrete nature of galaxies does not
allow for a bias that is both linear and deterministic. However, the
halo occupation statistics can in principle be such that the
bias is {\it linear} and {\it stochastic}, in which case
\begin{eqnarray}\label{linstoch}
\hatb = \wigb = b(M) = 1 & \;\;\;\;\;\; & \bvar = (1+\sigb^2)^{1/2} \nonumber\\
\sigb \ne 0 & \;\;\;\;\;\; & r = (1+\sigb^2)^{-1/2} \,,
\end{eqnarray}
so that $\bvar > 1$, while $r = 1/\bvar < 1$. In the case of
{\it non-linear, deterministic} biasing these relations reduce to
\begin{eqnarray}\label{nldet}
1 \ne \hatb \ne \wigb \ne 1 & \;\;\;\;\;\; & \bvar = \wigb \nonumber\\
\sigb = 0 & \;\;\;\;\;\; & r = \hatb/\wigb \ne 1 
\end{eqnarray}

\subsection{The Importance of Central and Satellite Galaxies}
\label{sec:censat}

An important aspect of halo occupation statistics is the split of
galaxies in two components: centrals and satellites. Centrals are
galaxies that reside at the center of their dark matter haloes, whereas
satellites orbit around centrals. As we will see below, this
distinction between centrals and satellites is the main cause for
non-linearity and scale-dependence in galaxy bias.  In order to gain
some insight into how centrals and satellites seperately contribute to
the stochasticity, we define their corresponding random halo biases
\begin{equation}
\varepsilon_\rmc \equiv N_\rmc - \langle N_\rmc|M \rangle \;\;\;\;\;\;
\varepsilon_\rms \equiv N_\rms - \langle N_\rms|M \rangle\,,
\end{equation}
where $N_\rmc$ and $N_\rms$ are the numbers of central and satellite
galaxies, respectively.  The halo stochasticity function for centrals
is given by
\begin{eqnarray}\label{stochcen}
\langle \varepsilon^2_\rmc | M \rangle & = & \sum_{N_\rmc = 0}^{1}
\left(N_\rmc - \langle N_\rmc|M\rangle \right)^2 \, P(N_\rmc|M) \nonumber \\
& = & \langle N_\rmc|M \rangle - \langle N_\rmc|M \rangle^2 \,.
\end{eqnarray}
Hence, we have that, as expected, central galaxies only contribute
stochasticity if $\langle N_\rmc|M \rangle < 1$.  On the other
  hand, if $\langle N_\rmc|M \rangle$ is unity, then
the occupation statistics of centrals are deterministic and $\langle
\varepsilon^2_\rmc|M \rangle = 0$. In the case of satellite galaxies
we have that
\begin{eqnarray}\label{stochsat}
\langle \varepsilon^2_\rms | M \rangle & = & \sum_{N_\rms = 0}^{\infty}
\left(N_\rms - \langle N_\rms|M\rangle \right)^2 \, P(N_\rms|M) \nonumber \\
& = & \langle N^2_\rms|M \rangle - \langle N_\rms|M \rangle^2 \,.
\end{eqnarray}
Introducing the {\it Poisson function}
\begin{equation}\label{Poipardef}
\beta(M) \equiv {\langle N_\rms (N_\rms - 1)|M \rangle \over
\langle N_\rms|M \rangle^2}\,,
\end{equation}
which is equal to unity if $P(N_\rms|M)$ is given by a Poisson
distribution, we can rewrite Eq.~(\ref{stochsat}) as
\begin{equation}\label{varsat}
\langle \varepsilon^2_\rms | M \rangle = \left[\beta(M) - 1\right] \,
\langle N_\rms|M \rangle^2 + \langle N_\rms|M \rangle\,.
\end{equation}
This makes it explicit that $\langle \varepsilon^2_\rms | M \rangle =
\langle N_\rms|M \rangle$ if the occupation statistics of satellite
galaxies obey Poisson statistics. Finally, the halo stochasticity
function for all galaxies (centrals and satellites combined) can be
written as
\begin{eqnarray}\label{varall}
\langle \varepsilon^2 | M \rangle & = & \langle N^2|M \rangle -
\langle N|M \rangle^2 \nonumber \\
& = & \langle \varepsilon^2_\rmc | M \rangle + 
\langle \varepsilon^2_\rms | M \rangle + \nonumber \\
& & 2 \left\{\langle N_\rmc\,N_\rms|M\rangle - \langle N_\rmc|M\rangle \;
\langle N_\rms|M \rangle \right\}\,.
\end{eqnarray}
where we have used that $N = N_\rmc + N_\rms$. Hence, as long as
$N_\rmc$ and $N_\rms$ are independent random variables, we have that
$\langle \varepsilon^2 | M \rangle$ is simply the sum of the halo
stochasticity functions for centrals and satellites.

\section{A Realistic Example}
\label{sec:example}

What are the typical values of $\hatb$, $\wigb$, $\sigb$, $\bvar$ and
$r$? In order to answer this question, 
we use a realistic model for the halo
occupation statistics, as described by the Conditional Luminosity
Function (CLF; \citealt{YMB03}). The CLF, $\Phi(L|M)$, specifies the
number of galaxies of luminosity $L$ that, on average, reside in a
halo of mass $M$. Following \citet{2005ApJ...627L..89C} and
\citet{YMB08} we split the CLF in a central and satellite component;
\begin{equation}\label{CLFsplit}
\Phi(L|M) = \Phi_\rmc(L|M) + \Phi_\rms(L|M)\,.
\end{equation}
We use the CLF parameterization of \citet{YMB08}, inferred from a
large galaxy group catalogue (\citealt{Y07}) extracted from the SDSS
Data Release 4 (\citealt{2006ApJS..162...38A}). In particular, the CLF
of central galaxies is modelled as a log-normal,
\begin{equation}\label{phi_c}
\Phi_\rmc(L|M) = {1 \over {\sqrt{2\pi} \, {\rm ln}(10)\, \sigma_\rmc} L} 
{\rm exp}\left[- { {(\log L  -\log L_\rmc )^2 } \over 2\,\sigma_\rmc^2} \right]\,
\,,
\end{equation}
and the satellite term as a modified Schechter function,
\begin{equation}\label{phi_s}
\Phi_\rms(L|M) = { \phi^*_\rms \over L^*_\rms}\,
\left({L\over L^*_\rms}\right)^{\alpha_\rms} \,
{\rm exp} \left[- \left ({L\over L^*_\rms}\right )^2 \right] 
\,.
\end{equation}
Note that $L_\rmc$, $\sigma_\rmc$, $\phi^*_\rms$, $\alpha_\rms$ and
$L^*_\rms$ are, in principle, all functions of halo mass $M$. As our
fiducial model, we adopt the specific CLF model of
\citet{2009MNRAS.394..929C},  with both $\sigma_\rmc$ and
  $\alpha_\rms$ assumed to be independent of halo mass. This model is
in excellent agreement with the observed abundances, clustering, and
galaxy-galaxy lensing properties of galaxies in the SDSS DR4 (see
\citealt{2009MNRAS.394..929C}) as well as with satellite kinematics
(\citealt{2009MNRAS.392..801M}). In other words, this particular CLF
provides a realistic and accurate description of the halo occupation
statistics, at least for the cosmology adopted here.

From the CLF it is straightforward to compute the halo occupation
numbers.  For example, the average number of galaxies with
luminosities in the range $L_1 \leq L \leq L_2$ is simply given by
\begin{equation}\label{HODfromCLF}
\langle N|M \rangle = \int_{L_1}^{L_2} \Phi(L|M) \, \rmd L\,.
\end{equation}
The CLF, however, only specifies the first moment of the halo
occupation distribution $P(N|M)$.  For central galaxies, $\langle
N^2_\rmc|M \rangle = \langle N_\rmc|M \rangle$, which simply follows
from the fact that $N_\rmc$ is either zero or unity.  For satellite
galaxies, we use that 
\begin{equation}
\langle N_\rms^2|M \rangle = \beta(M) \langle N_\rms|M \rangle^2 + 
\langle N_\rms|M \rangle
\end{equation}
where $\beta(M)$ is the Poisson function [Eq.~(\ref{Poipardef})]. In
what follows we limit ourselves to cases in which $\beta(M)$ is
independent of halo mass, i.e., $\beta(M) = \beta$, and we treat
$\beta$ as a free parameter. In our fiducial model we set $\beta=1$,
so that $P(N_\rms|M)$ is a Poisson distribution.
\begin{figure*}
\psfig{figure=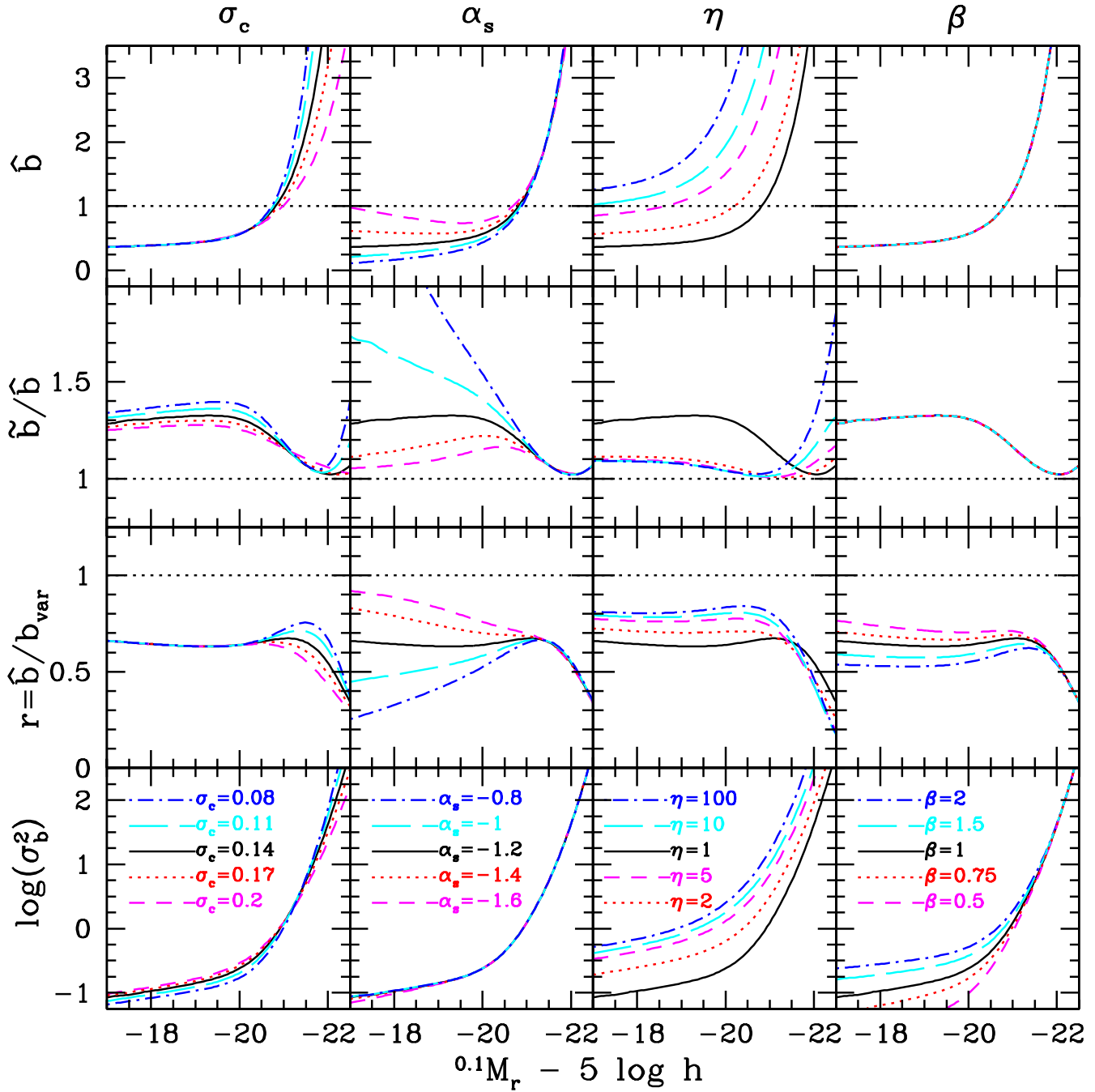,width=\hdsize}
\caption{The dependence of galaxy bias on halo occupation statistics.
  From top to bottom, the various panels plot $\hatb$, the
  non-linearity parameter $\wigb/\hatb$, the cross correlation
  coefficient $r$, and the stochasticity parameter $\log(\sigb^2)$
  always for galaxies with luminosities in the range
  $[L_1,2.5L_1]$. Results are plotted as function of $L_1$, expressed
  as an $r$-band magnitude that has been $K$+$E$ corrected to
  $z=0.1$. From left to right, we each time vary only one parameter
  (listed at the top) with respect to our fiducial model, which is
  indicated by the black, solid line. The horizontal, dotted line
  corresponds to linear, deterministic biasing.  
  See eq.(\ref{phi_c}), (\ref{phi_s}),(\ref{eq:eta}), and (\ref{Poipardef}) 
  for the definition of $\sigma_{\rm c}, \alpha_{\rm s}, \eta,$ and $\beta$, 
  respectively.
  }
\label{fig:CLFbias}
\end{figure*}

\begin{figure*}
\psfig{figure=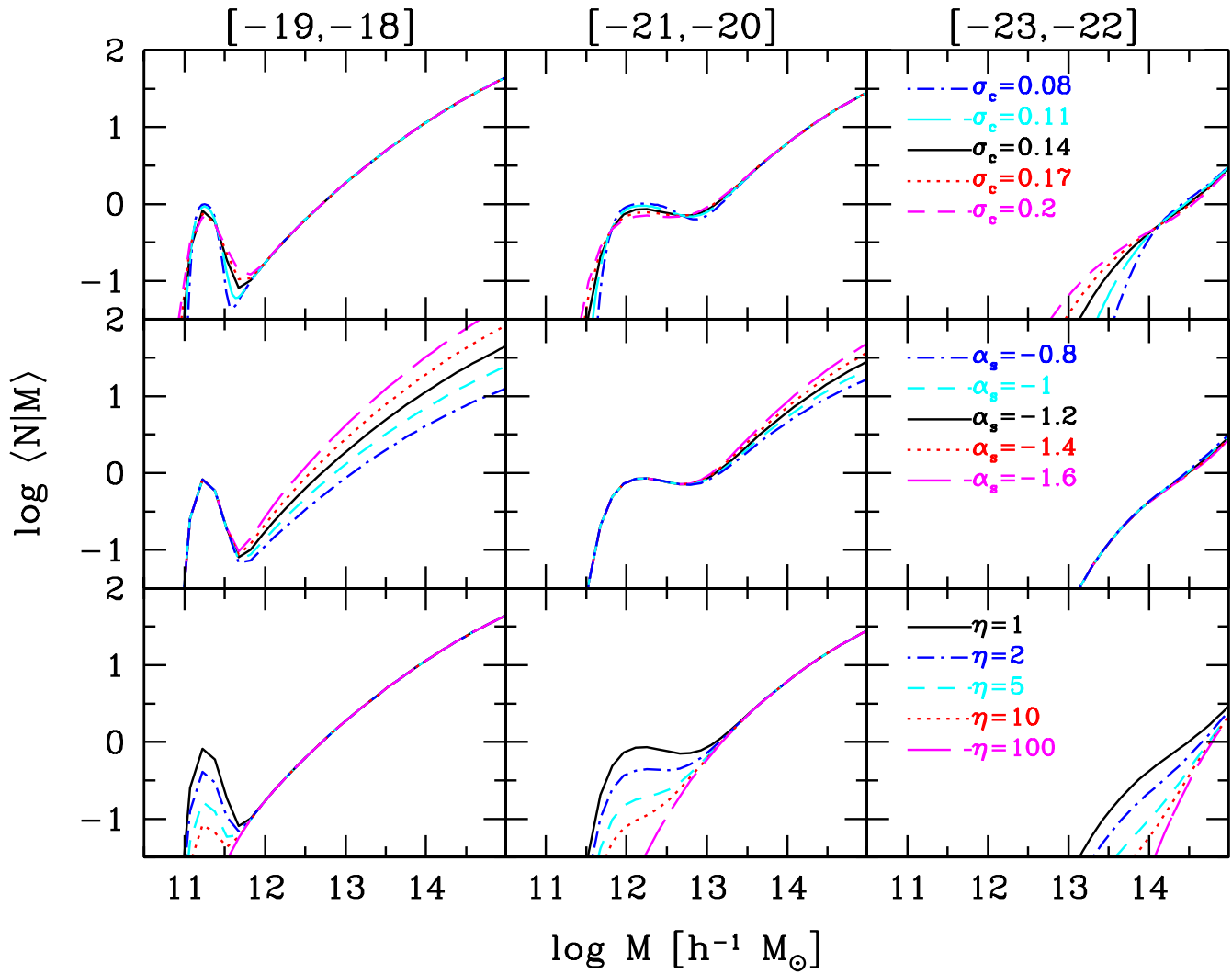,width=\hdsize}
\caption{The halo occupation statistics, $\langle N|M \rangle$, as
  function of halo mass, $M$, for three different magnitude bins
  (different columns, as indicated at the top). From top to bottom we
  vary $\sigma_\rmc$, $\alpha_\rms$ and $\eta$ with respect to our
  fiducial model, which is always indicated as a black, solid
  curve. These are the same HODs as used in
  Fig.~\ref{fig:CLFbias}. Note how increasing $\sigma_\rmc$ broadens
  the contribution of centrals, how increasing $\alpha_\rmc$ reduces
  the slope $\rmd \log \langle N|M \rangle/\rmd\log M$, and how
  increasing $\eta$ suppresses the contribution of centrals.}
\label{fig:HODs}
\end{figure*}

Fig.~\ref{fig:CLFbias} shows the parameters $\hatb$ (upper row),
$\wigb/\hatb$ (second row), $r$ (third row), and $\log(\sigb^2)$
(bottom row) as functions of $L_1$ (expressed in $^{0.1}M_r - 5\log
h$, which is the $r$-band magnitude $K$+$E$ corrected to $z=0.1$).
Throughout we have adopted luminosity bins of one magnitude width, so
that $L_2 = 2.5 L_1$; hence, the value of $\hatb$ at $^{0.1}M_r -
5\log h = -20$ indicates the first moment of the mean biasing
function, $b(M)$, for galaxies with $-21 \leq {^{0.1}M}_r - 5\log h
\leq -20$, etc. The solid curves in Fig.~\ref{fig:CLFbias} show the
results for our fiducial model ($\sigma_\rmc = 0.14$; $\alpha_\rms =
-1.2$; $\eta = 1.0$ and $\beta = 1.0$), while the different columns
show models in which we vary only one of the CLF parameters, as
indicated.  The parameter $\eta$ in the third column is defined by
\begin{equation}\label{eq:eta}
\Phi(L|M) = \Phi_\rmc(L|M)/\eta + \Phi_\rms(L|M) \, , 
\end{equation}
and is used to artificially reduce the contribution of centrals,
simply by increasing $\eta$ with respect to its fiducial value.  To
guide the discussion, Fig.~\ref{fig:HODs} shows the halo occupation
distributions (HODs) corresponding to the various models shown in
Fig.~\ref{fig:CLFbias}.  Note how increasing $\sigma_\rmc$ broadens
the contribution of centrals, how increasing $\alpha_\rmc$ reduces the
slope $\rmd \log \langle N|M \rangle/\rmd\log M$, and how increasing
$\eta$ suppresses the contribution of centrals.

Starting with the upper panels of Figure~\ref{fig:CLFbias}, 
we see that the fiducial model predicts a $\hatb(L_1)$ that strongly 
increases with luminosity.  In order to understand this behavior, 
we use Eqs.~(\ref{bMdef}) and~(\ref{moments}) to write
\begin{equation}\label{bhatmod}
\hatb = {\averrho \over \averng} \, {\int \langle N|M \rangle \, M 
\, n(M) \, \rmd M \over \int M^2 \, n(M) \, \rmd M} 
\end{equation}
Hence, $\hatb=1$ if $\langle N|M \rangle = (\averng/\averrho) \, M$,
which corresponds to $b(M) = 1$. As evident from
Fig.~\ref{fig:HODs}, for brighter bins, $\langle N|M \rangle$ cuts off
at higher $M$ (i.e., bright galaxies can only reside in massive
haloes), and this cut-off causes $\hatb > 1$. For fainter galaxies,
the contribution of central galaxies in the HOD becomes more
pronounced, resulting in a `boost' of $\langle N|M \rangle$ at
relatively low $M$ (see left-hand panels of Fig.~\ref{fig:HODs}),
which in turn results in $\hatb < 1$. This is also evident from the
scaling with $\eta$: increasing $\eta$ suppresses the relative
contribution of centrals, which in turn causes an increase in
$\hatb$. Changing the Poisson parameter only changes the scatter in
the number of satellites, and therefore has no impact on $\hatb$ (or
$\wigb$), while changes in $\sigma_\rmc$ or $\alpha_\rms$ has only a
mild impact on $\hatb$ for reasons that are easily understood from an
examination of Eq.~(\ref{bhatmod}) and Fig.~\ref{fig:HODs}.  The main
message here is that realistic HODs differ strongly from the simple
scaling $\langle N|M \rangle \propto M$, such that $\hatb < 1$ ($\hatb
> 1$) for faint (bright) galaxies.

The second row of panels in Fig.~\ref{fig:CLFbias} shows that the
`normalized' non-linearity parameter $\wigb/\hatb$ also differs
markedly from unity for our fiducial model. As for $\hatb$, this
parameter is also equal to unity if bias is linear (i.e., if $\langle
N|M \rangle = (\averng/\averrho) \, M$). Since this is not the case
for realistic HODs, both $\wigb$ and $\hatb$ will in general differ
from unity.  At the faint-end, $\wigb/\hatb$ is extremely sensitive to
the parameter $\alpha_\rms$. This is easy to understand; as is evident
from Fig.~\ref{fig:HODs}, the parameter $\alpha_\rms$ controls the
slope of $\langle N|M \rangle$ at the massive end, especially for
fainter galaxies, for which the satellite fraction is larger. Models
for which the slope $\rmd \log \langle N|M \rangle/\rmd\log M$
deviates more from unity are more non-linear.  In other words,
$\wigb/\hatb$ is simply a measure for how much $\langle N|M \rangle$
differs from the linear relation $\langle N|M \rangle \propto M$.

The third row of panels in Fig.~\ref{fig:CLFbias} shows the
cross-correlation coefficient $r$. In all cases shown, and for
all luminosities, the CLF indicates that $r < 1$. Writing that
\begin{equation}\label{CCCrecast}
r = {\hatb\over\wigb} \, \left[ 1 + 
\left({\sigb\over\wigb}\right)^2\right]^{-1/2}\,,
\end{equation}
we immediately see that $r \leq \hatb/\wigb$ (where the equality
corresponds to deterministic biasing). Hence, the fact that $r < 1$
simply reflects the fact that, for realistic halo occupation
statistics, the non-linearity parameter $\wigb/\hatb > 1$. The decline
of $r$ at the bright-end is a reflection of stochasticity becoming
more and more important for brighter galaxies. This is evident from
the bottom panels of Fig.~\ref{fig:CLFbias}, which show that $\sigb$
increases very strongly with luminosity. In order to understand this
behavior, it is useful to rewrite Eq.~(\ref{stochpardef}) using
Eqs.~(\ref{varsat})-(\ref{varall}) as
\begin{equation}\label{sigbrecast}
\sigb = {1 \over \calM_2}  \left[{1\over\averng} + \calI(\beta)\right]^{1/2}\,,
\end{equation}
where
\begin{equation}
\calM_2 \equiv \left[\int \left({M\over\averrho}\right)^2 \, n(M) \, 
\rmd M\right]^{1/2}
\end{equation}
is some constant, we have assumed that $N_\rmc$ and $N_\rms$ are
independent random variables, and
\begin{eqnarray}\label{calIbeta}
\calI(\beta) & \equiv & (\beta-1) \, f^2_\rms \, \int
{\langle N_\rms|M \rangle^2 \over \averns^2} \, n(M) \, \rmd M - \nonumber\\
& & f^2_\rmc \, \int {\langle N_\rmc|M \rangle^2\over \avernc^2} 
\, n(M) \, \rmd M\,.
\end{eqnarray}
Here $f_\rmc = \bar{n}_\rmc/\bar{n}_\rmg$ and $f_\rms =
\bar{n}_\rms/\bar{n}_\rmg = 1 - f_\rmc$ are the central and satellite
fractions, respectively, and the average number densities $\averng$,
$\avernc$ and $\averns$ follow from
\begin{equation}\label{averng}
\avernx = \int \langle N_\rmx|M \rangle \, n(M) \, \rmd M\,,
\end{equation}
where `x' stands for `g' (for galaxies), `c' (for centrals) or `s'
(for satellites). The first term of Eq.~(\ref{sigbrecast}) indicates
the contribution to $\sigb$ due to shot-noise, i.e., the finite number
of galaxies (per halo). This term dominates when the number density of
galaxies is small (i.e., for bright galaxies), in which case $\sigb
\propto \averng^{-1/2}$. It is this shot-noise that is responsible for
driving $r \rightarrow 0$ at the bright end.  The second term of
Eq.~(\ref{sigbrecast}) describes the contribution to $\sigb$ due to
specific aspects of the halo occupation statistics, as described by
the function $\calI(\beta)$. This function, which is independent of
$\averng$, consists of two terms describing the contributions due to
the possible non-Poissonian nature of satellite galaxies (i.e., if
$\beta \ne 1$) and due to scatter in the occupation statistics of
centrals (i.e., a non-zero $\sigma_\rmc$). Note that the first term of
$\calI$ is zero for our fiducial model, which has $\beta=1$. This
explains the insensitivity\footnote{Since
  the number density of galaxies is dominated by centrals, changes in
  the number density of satellites due to changes in $\alpha_\rms$
  have almost no impact on $\sigb$.} to changes in $\alpha_\rms$. 
  Changes in $\sigma_\rmc$ only
have a very mild impact on the stochasticity, while increasing
(decreasing) $\beta$ only has a significant impact for faint galaxies,
simply because the $1/\averng$ shot-noise term dominates for bright
galaxies. Finally, the increase of $\sigb$ with increasing $\eta$ is
simply a reflection of the fact that an increase in $\eta$ reduces the
number density of (central) galaxies.

To summarize, realistic halo occupation models predict a galaxy bias
that is strongly non-linear, indicating that realistic models do not
scale as $\langle N|M \rangle \propto M$. This is mainly a consequence
of central galaxies, which dominate the number density and for which
$\langle N_\rmc|M \rangle$ never resembles a power-law.  However, even
for satellite galaxies it is important to realize that $\langle
N_\rms|M \rangle$ never follows a pure power law; even if $\langle
N_\rms|M \rangle \propto M$ at the massive end, there will be a
cut-off at low $M$, reflecting that galaxies of a given luminosity (or
stellar mass) require a minimum mass for their host halo. Such a
cut-off by itself is already sufficient to cause $\hatb$ and $\wigb$
to deviate significantly from unity. As for the stochasticity; this is
largely dominated by shot-noise, with halo-to-halo scatter, which
reflects the second moment of the halo occupation distribution
$P(N|M)$, only contributing significantly for fainter galaxies.

\section{Two-Point Statistics}
\label{sec:2pstat}

The bias parameters defined in \S\ref{sec:galbias} are quantities that
are averaged over dark matter haloes. Given that (virtually) all
galaxies are believed to reside in haloes, this is a logical and
intuitive way to define galaxy bias. However, observationally it is
far from trivial to actually measure these quantities from
data. After all, this requires an observational method to identify
individual dark matter haloes. In principle, this can be achieved
using gravitational lensing, but in practice this is only possible for
massive clusters. An alternative is to use a halo-based galaxy group
finder, such as the one developed by \citet{Yang2005}. However, this
method has the disadvantage that it uses galaxies to identify the dark
matter haloes and estimate their masses. Consequently, $N$ and $M$ are
not independent variables, which is likely to cause systematic
errors. For example, if $N$ is in one way or the other used to
estimate $M$, the halo-to-halo variance in $P(N|M)$, and hence the
amount of stochasticity, will be underestimated.

Therefore, when the goal is to put constraints on galaxy bias using
observational data, one requires another set of bias parameters that
do not suffer from these shortcomings. Such a set can be defined using
two-point statistics, such as the galaxy-galaxy and galaxy-matter
correlation functions, which can be reliably measured from large
galaxy surveys such as the SDSS. In addition, with the help of the
`halo model', which describes the dark matter density field in terms
of its halo building blocks (see e.g.,
\citealt{2002PhR...372....1C,MBW10}), one can analytically compute the
galaxy-galaxy correlation function from the same occupation
statistics, $P(N|M)$, required to compute $\hatb$, $\wigb$, $b_{\rm
  var}$, $r$ and $\sigb$.

Let us define the following two-point correlation functions:
\begin{eqnarray}\label{eq:corrdef}
\xi_{\rm gg}(r) &\equiv & \langle \delta_{\rm g}({\bf x}) 
\delta_{\rm g}({\bf x}+{\bf r}) \rangle \nonumber \\
\xi_{\rm mm}(r) &\equiv & \langle \delta_{\rm m}({\bf x})
\delta_{\rm m}({\bf x}+{\bf r}) \rangle \nonumber \\
\xi_{\rm gm}(r) &\equiv & \langle \delta_{\rm g}({\bf x})
\delta_{\rm m}({\bf x}+{\bf r}) \rangle\,,  
\end{eqnarray}
where $\langle ... \rangle$ represents an ensemble average, $r = |{\bf
  r}|$ is the distance between the two locations\footnote{Here we have
  made the standard assumption that the Universe is isotropic.}, and
the subscripts `gg', `gm', `mm' refer to `galaxy-galaxy',
`galaxy-matter', and `matter-matter', respectively. 

Using these two-point correlation functions, we now define three
functions that are sensitive to different aspects of galaxy bias:
the `classical' galaxy bias function,
\begin{equation}\label{eq:bgaldef}
b^{\rm 3D}_\rmg(r) \equiv \sqrt{\frac{\xi_{\rm gg}(r)}{\xi_{\rm mm}(r)}} \,,
\end{equation}
the galaxy-dark matter cross-correlation coefficient (hereafter CCC),
\begin{equation}\label{eq:CCCdef}
\calR^{\rm 3D}_{\rm gm}(r) \equiv \frac{\xi_{\rm gm}(r)}
{[\xi_{\rm gg}(r)*\xi_{\rm mm}(r)]^{1/2}}\,,
\end{equation}
(\citealt{1998ApJ...504..601P}), and their ratio,
\begin{equation}\label{eq:Gammadef}
\Gamma^{\rm 3D}_{\rm gm}(r) \equiv 
\frac{b^{\rm 3D}_{\rm g}(r)}{{\cal R}^{\rm 3D}_{\rm gm}(r)} = 
\frac{\xi_{\rm gg}(r)}{\xi_{\rm gm}(r)}\, .
\end{equation} 
The reason for introducing $\Gamma^{\rm 3D}_{\rm gm}$ is that,
contrary to $b^{\rm 3D}_{\rm gm}$ and $\calR^{\rm 3D}_{\rm gm}$ it is
independent of the matter-matter correlation function, which makes it
easier to measure observationally (see \citealt{2004AJ....127.2544S}).
In what follows we shall loosely refer to these functions as `bias
functions', and to their $r$-dependence as `scale-dependence'.

Using the ergodic principle, the ensemble average, $\langle
... \rangle$, can be written as a volume average, which in turn, under
the assumption that all dark matter is bound in virialized dark matter
haloes, is equal to an average over dark matter haloes (see
\citealt{MBW10}). Hence, similar to the bias parameters defined in
\S\ref{sec:galbias}, the bias functions $b^{\rm 3D}_\rmg(r)$,
$\calR^{\rm 3D}_{\rm gm}(r)$, and $\Gamma^{\rm 3D}_{\rm gm}(r)$ are
also `defined' as halo averaged quantities. The advantage of defining
bias functions via two-point statistics, however, is that they can be
measured without the need to identify individual dark matter haloes.
Finally, we emphasize that the CCC is not restricted to $\vert
\calR(r) \vert \leq 1$ when computed with the analytical halo model
because it intrinsically subtracts out the shot-noise term in the
galaxy correlation (see e.g.,
\citealt{1999MNRAS.304..767S,2000MNRAS.318..203S}).
\begin{figure*}
\psfig{figure=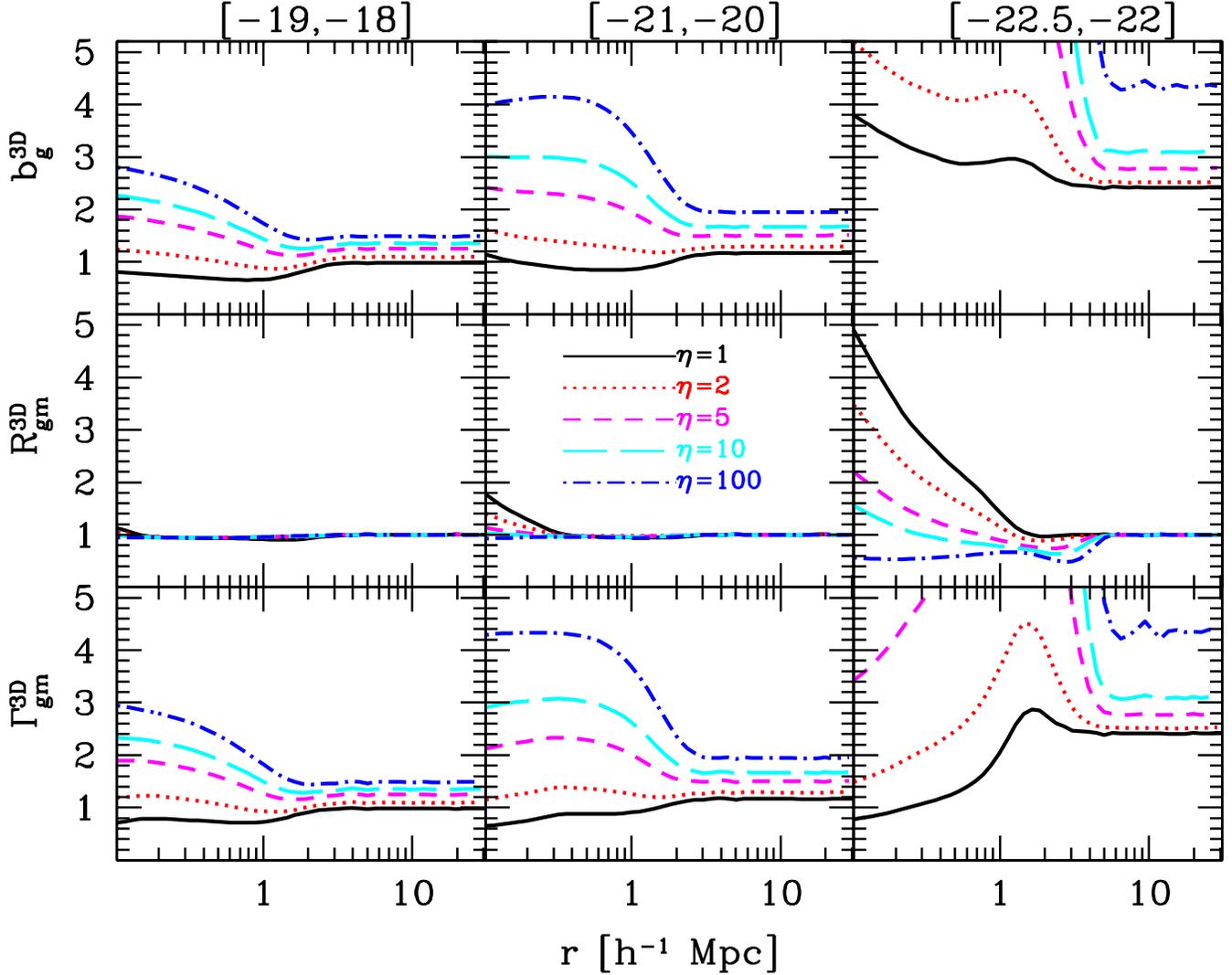,width=\hdsize}
\caption{
Scale dependence of the galaxy bias functions $b^{\rm 3D}_{\rm g}$,
 ${\cal R}^{\rm 3D}_{\rm gm}$, and $\Gamma^{\rm 3D}_{\rm gm}$ for 
 three luminosity bins (indicated at the top of every column). The reference
 model is indicated with the black solid lines, whereas other lines refer to 
 a suppression of the central term of the  HOD by a factor $1/\eta$ 
 (see eq.~[\ref{eq:eta}] and discussion in \S3).}
\label{fig:fig1b}
\end{figure*}

\subsection{Analytical Model}
\label{sec:analmodel}

As we shall see below, all these three bias functions can be computed
analytically from the halo occupation distribution $P(N|M)$, making
them the natural extension of the bias parameters defined in
\S~\ref{sec:galbias} to the two-point regime. We address the
observational perspective of these bias functions in
\S\ref{sec:observations}. In this section we focus on investigating
what realistic models for halo occupation statistics predict for (the
scale dependence of) $b^{\rm 3D}_\rmg$, $\calR^{\rm 3D}_{\rm gm}$ and
$\Gamma^{\rm 3D}_{\rm gm}$.

In what follows we describe the two-point statistics in Fourier space,
using power-spectra rather than two-point correlation functions. This
has the single advantage that equations that involve convolutions in
real-space can now be written in a more compact form.  Throughout we
assume that dark matter haloes are spherically symmetric and have a
density profile, $\rho_{\rm m}(r|M) = M \, u_\rmh(r|M)$, that depends
only on their mass, $M$. Note that $\int u_\rmh(\bx|M,z) \, \rmd^3 \bx
= 1$. Similarly, we assume that satellite galaxies in haloes of mass
$M$ follow a  spherical number density distribution $n_\rms(r|M)
= N_\rms \, u_\rms(r|M)$, while central galaxies always have
$r=0$. Since centrals and satellites are distributed differently, we
write the galaxy-galaxy power spectrum as
\begin{equation}\label{Pkgalsplit}
P_{\rm gg}(k) = f^2_\rmc P_{\rm cc}(k) + 
2 f_\rmc f_\rms P_{\rm cs}(k) + f^2_\rms P_{\rm ss}(k)\,,
\end{equation}
while the galaxy-dark matter cross power spectrum is given by
\begin{equation}\label{Pkgaldmsplit}
P_{\rm gm}(k) = f_\rmc P_{\rm cm}(k) + f_\rms P_{\rm sm}(k)\,.
\end{equation}
In addition, it is common practice to split two-point statistics into
a 1-halo term (both points are located in the same halo) and a 2-halo
term (the two points are located in different haloes). Following
\citealt{2002PhR...372....1C} and \citealt{MBW10}, the 1-halo terms
are
\begin{equation}
P^{\rm 1h}_{\rm cc}(k) = {1 \over \bar{n}_\rmc}\,,
\end{equation}
\begin{equation}
P^{\rm 1h}_{\rm ss}(k) = \beta \int \calH_{\rms}^2(M) \, n(M) \, \rmd M\,,
\end{equation}
and all other terms are given by
\begin{equation}
P^{\rm 1h}_{\rm xy}(k) = \int \calH_{\rm x}(M) \, \calH_{\rm y}(M) \, n(M) 
\, \rmd M\,.
\end{equation}
Here `x' and `y' are either `c' (for central), `s' (for satellite), or
`m' (for matter), and we have defined
\begin{equation}
\calH_\rmm(M) = {M \over \averrho} \,  \tilde{u}_\rmh(k|M)\,,
\end{equation}
\begin{equation}
\calH_\rmc(M) = {\langle N_\rmc|M \rangle \over \bar{n}_{\rmc}} \,,
\end{equation}
and
\begin{equation}
\calH_\rms(M) = {\langle N_\rms|M \rangle \over \bar{n}_{\rms}} \,  
\tilde{u}_\rms(k|M)\,,
\end{equation}
with $\tilde{u}_\rmh(k|M)$ and $\tilde{u}_\rms(k|M)$ the Fourier
transforms of the halo density profile and the satellite number
density profile, respectively, both normalized to unity. The various
2-halo terms are given by
\begin{eqnarray}\label{P2hcc}
\lefteqn{P^{\rm 2h}_{\rmx\rmy}(k) = P_{\rm lin}(k) \,  \int \rmd
  M_1 \, \calH_\rmx(M_1) \, b_{\rm h}(M_1)\, n(M_1)} \nonumber \\ 
& & \int \rmd M_2 \, \calH_\rmy(M_2) \, b_{\rm h}(M_2)\, n(M_2) \,,
\end{eqnarray}
where $P_{\rm lin}(k)$ is the linear power spectrum and $b_\rmh(M,z)$
is the halo bias function (e.g., \citealt{1996MNRAS.282..347M}). 
Note that in this formalism, the matter-matter power spectrum simply reads
$P_{\rm mm}(k) = P^{\rm 1h}_{\rm mm}(k) + P^{\rm 2h}_{\rm mm}(k) \, .$

The two-point correlation functions corresponding to these power-spectra
are obtained by simple Fourier transformation:
\begin{equation}\label{xiFTfromPK}
\xi_{\rm xy}(r) = {1 \over 2 \pi^2} \int_0^{\infty} P_{\rm xy}(k) \,
{\sin kr \over kr} \, k^2 \, \rmd k\,, 
\end{equation}
Throughout this paper we adopt the halo mass functions and halo bias
functions of \citet{2008ApJ...688..709T} and
\citet{2010ApJ...724..878T}, respectively.

We caution the reader that this particular implementation of the halo
model is fairly simplified.  In particular, it ignores two important
effects: scale dependence of the halo bias function and halo-exclusion
(i.e., the fact that the spatial distribution of dark matter haloes is
mutually exclusive). As discussed in \citet{2005ApJ...631...41T}, both
effects are important on intermediate scales (in the 1-halo to 2-halo
transition region). Indeed, demonstrated in van den Bosch \etal (2012,
in preparation), ignoring these effects can cause systematic errors in
the two-point correlation functions on scales of $\sim 1-2 h^{-1}\Mpc$
that are as large as 50 percent. However, detailed tests have shown
that the bias functions (\ref{eq:bgaldef})-(\ref{eq:Gammadef}), which
are defined in terms of ratios of these correlation functions, are
much more accurate (with typical errors $\lta 10$ percent); i.e., to
first order, by taking ratios, one is less sensitive to uncertainties
in the halo model. This is why it is advantageous to use the bias
functions, rather than the two-point correlation functions themselves,
when trying to constrain particular aspects of galaxy bias. It is the
purpose of this paper to explore how the (scale-dependence) of the
bias functions depend on various properties related to halo occupation
statistics.
\begin{figure*}
\psfig{figure=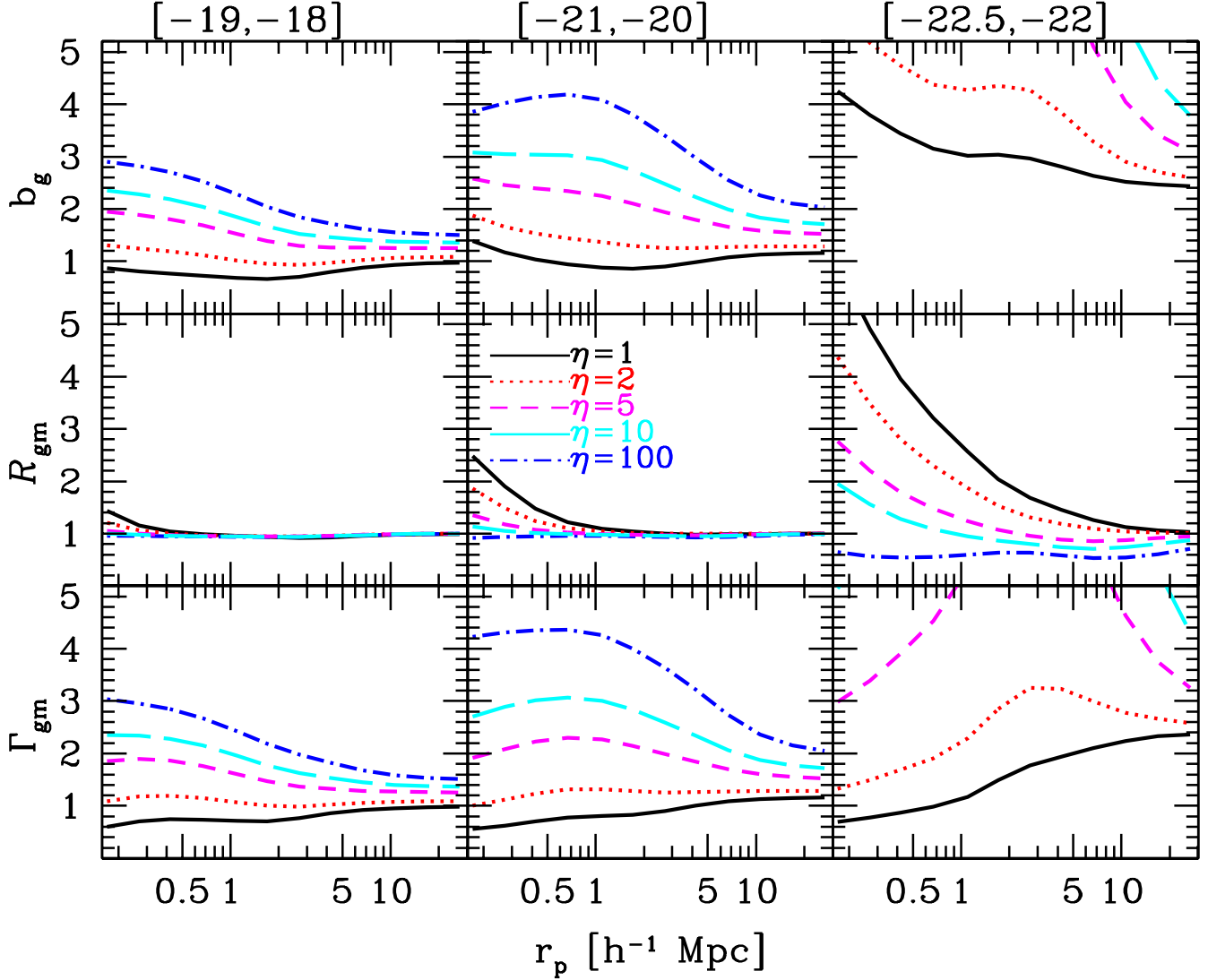,width=\hdsize}
\caption{ Scale dependence of the galaxy bias functions $b_{\rm g}$,
  ${\cal R}_{\rm gm}$, and $\Gamma_{\rm gm}$ for three luminosity bins
  (indicated at the top of every column). The reference model is
  indicated with the black solid lines, whereas other lines refer to a
  suppression of the central term of the  HOD by a factor
  $1/\eta$ (see discussion in \S3).  }
\label{fig:fig4}
\end{figure*}

Before showing predictions based on a specific HOD, we can gain some
insight from a closer examination of the above equations. In
particular, it can easily be seen that galaxies are only unbiased with
respect to the dark matter, at all scales (i.e., $b^{\rm 3D}_{\rm g} =
1$), if and only if the following conditions are satisfied
\begin{enumerate}
\item $\eta = \infty$, i.e., there are no central galaxies
\item $\beta=1$, i.e., the occupation number of satellite galaxies
  obeys Poisson statistics
\item $u_\rms(k|M) = u_\rmh(k|M)$, i.e., the normalized number density
  profile of satellite galaxies in dark matter haloes is identical to
  that of dark matter particles
\item $\langle N_\rms|M \rangle = {{\bar n}_\rms\over\rho} M$, i.e.,
  the occupation number of satellites is directly proportional to halo
  mass
\end{enumerate}
Under these conditions one also has that $\calR^{\rm 3D}_{\rm gm} =
\Gamma^{\rm 3D}_{\rm gm} = 1$. If all the above conditions are
satisfied except that there is a non-negligble fraction of centrals
(i.e., $\eta$ is finite), one expects strong scale-dependence on small
scales due to the fact that the location of central galaxies within
their dark matter haloes is strongly biased.  If $\beta \ne 1$, one
still maintains $b_\rmg^{\rm 3D} = 1$ on large scales ($r \gg
r_{12}$), but this will transit to $b_\rmg^{\rm 3D} = \beta$ for $r
\ll r_{12}$ (if $\eta = \infty$). Here $r_{12}$ is the 1-halo to
2-halo transition region, which is roughly equal to the virial radius
of the characteristic halo hosting the galaxies in question. This
comes about because the Poisson parameter $\beta$ only enters in the
1-halo satellite-satellite term.  If $\tilde{u}_\rms(k|M) \ne
\tilde{u}_\rmh(k|M)$, once again this will manifest itself as a
transition from $b_\rmg^{\rm 3D} = 1$ at $r \gg r_{12}$ to
$b_\rmg^{\rm 3D} \ne 1$ at $r \ll r_{12}$ simply because
$\tilde{u}_\rms(k|M) = \tilde{u}_\rmh(k|M) = 1$ on large scales (i.e.,
for $k \ll 1/r_{12}$). If $\langle N_\rms|M \rangle$ is not
proportional to $M$, which as we have seen in \S\ref{sec:example} is
never expected to be the case for realistic HODs, one also expects a
transition around $r_{12}$. However, in this case $b_\rmg^{\rm 3D}$ is
not expected to be equal to unity at either small or large scales. The
exact values of $b_\rmg^{\rm 3D}$ depend on how exactly the satellite
occupation numbers deviate from linearity, but will be different on
small and large scales mainly because the 2-halo term weights $\langle
N_\rms|M \rangle$ with the halo bias $b_\rmh(M)$, whereas the 1-halo
term doesn't. In fact, on large scales, where $\tilde{u}_\rms(k|M) =
\tilde{u}_\rmh(k|M) = 1$ and the matter power spectrum is still in the
linear regime, it is straightforward to show that
\begin{equation}\label{bglargescale}
b^{\rm 3D}_\rmg = 
{\langle M\,b(M)\,b_\rmh(M) \rangle \over \langle M \rangle}\,,
\end{equation}
where $\langle ... \rangle$ is the average over dark matter haloes
given by Eq.~(\ref{haloaver}) and $b(M)$ is the mean biasing function
of Eq.~(\ref{bMdef}). This immediately shows that the large scale bias
$b_\rmg$ defined via the correlation functions cannot, in general, be
expressed in terms of the moments $\hatb$ and $\wigb$ of $b(M)$
(cf. Eq.~[\ref{moments}]).  Only in the case of linear biasing we have
that $b^{\rm 3D}_\rmg = b(M) = \hatb = \wigb = 1$. To summarize, based
on all these considerations, one expects scale independence on large
scales, at a value that depends on the details of the HOD, a sudden
transition at around the 1-halo to 2-halo transition scale, and scale
dependence on small scales due to the dominance of central
galaxies. Finally, it is worth pointing out that on the large scales
where Eq.~(\ref{bglargescale}) is valid, one always expects that
$\calR^{\rm 3D}_{\rm gm}=1$ and $\Gamma^{\rm 3D}_{\rm gm} = b^{\rm
  3D}_\rmg$.

Figure 3 shows the scale dependence of the biasing functions defined
in Eqs.~(\ref{eq:bgaldef})-(\ref{eq:Gammadef}) computed using the
analytic model outlined above and with the same halo occupation
statistics as in \S\ref{sec:example}.  The three columns refer to
three luminosity bins (expressed in r-band magnitude). Beside the
reference model (black solid lines), we show four additional model
predictions in which the contribution from central galaxies is
increasingly suppressed as the parameter $\eta$ changes from unity to
100 (see discusssion in \S3). As is evident from the upper panels, the
galaxy bias, $b^{\rm 3D}_{\rm g}$, exactly reveals the behaviour
expected based on the discussion above: on large scales the bias is
scale-independent, there is sudden transition around the 1-halo to
2-halo transition scale (which is larger for brighter galaxies, since
these reside in more massive, and therefore more extended haloes), and
there is significant scale dependence on small scales.  Note also that
the large-scale bias is larger for brighter galaxies. This is
consistent with observations (e.g., \citealt{2000A&A...355....1G, N01,
  N02, 2005ApJ...630....1Z, 2007ApJ...664..608W,
  2011ApJ...736...59Z}), and is a manifestation of the fact that
brighter galaxies reside in more massive haloes, which are more
strongly clustered (\citealt{1996MNRAS.282..347M}). The suppression of
the contribution from central galaxies (i.e., increasing $\eta$) has
the effect of increasing the value of the bias on large scales and it
also affects the scale dependence of the bias on small scales.  The
effect is larger for brighter galaxies which is a direct consequence
of the fact that, for realistic HODs the fraction of galaxies which
are centrals is an increasing function of luminosity (e.g.,
\citealt{2006MNRAS.368..715M,vdB07,2009MNRAS.394..929C}).

As expected, the CCC, $\calR_{\rm gm}^{\rm 3D}$, shown in the panels
in the middle row, is unity on large scales for all three luminosity
bins, and independent of the value of $\eta$.  On small scales,
however, $\calR^{\rm 3D}_{\rm gm} > 1$. This scale dependence is
mainly due to the central galaxies being located at the centers of
their dark matter haloes.  Indeed, suppressing the contribution of
central galaxies (i.e., increasing $\eta$) results in a CCC that is
closer to unity on small scales.  Overall, the scale dependence of
$\calR^{\rm 3D}_{\rm gm}$ is more pronounced for brighter
galaxies. This is a consequence of the fact that brighter galaxies
reside in more massive, and therefore more extended, haloes. Note that
our model, which is based on a realistic HOD that is in excellent
agreement with a wide range of data, predicts that for galaxies with
magnitudes in the range $-18 \geq ^{0.1}M_r - 5\log h \geq -19$ 
and  $-20 \geq ^{0.1}M_r - 5\log h \geq -21$
the CCC is very close to unity 
on scales above $r\gsim0.2 h^{-1}$ Mpc and $r\gsim0.4 h^{-1}$ Mpc, respectively. 
As we will see below, this is actually a very robust prediction, and
indicates that if suppression of scale-dependence is important, it is
in general advantageous to use fainter galaxies (but see
\S\ref{sec:Reyes}). For the brightest bin, our model predicts that 
simply reducing the contribution from the central galaxies does not 
lead to $\calR^{\rm 3D}_{\rm gm}=1$ over the scale probed here. 
This is due to the fact that a realistic HOD does not have 
$\langle N_\rms|M \rangle \propto M$ and even more important 
bright satellite galaxies only form above a large halo mass. 
We have tested that artificially imposing $\langle N_\rms|M \rangle \propto M$
and no low-mass cut off yields $\calR^{\rm 3D}_{\rm gm}=1$
at all scales of interested here also in this brightest bin.

Finally, because of its definition, the scale dependence of the bias
function $\Gamma^{\rm 3D}_{\rm gm}$, shown in the lower panels, is
easily understood from a combination of the effects on both $b^{\rm
  3D}_{\rm gm}$ and ${\cal R}^{\rm 3D}_{\rm gm}$. Our fiducial model
predicts that $\Gamma^{\rm 3D}_{\rm gm}$ is scale-independent on large
scales, and decreases with decreasing radius on small scales where the
1-halo term of the two-point correlation functions dominates. Overall,
the scale-dependence of $\Gamma^{\rm 3D}_{\rm gm}$ is predicted to be
larger for brighter galaxies.

The results in Fig.~3 indicate that our analytical model, which is
based on a realistic HOD, makes some very specific predictions
regarding the scale dependence of the bias
functions~(\ref{eq:bgaldef})-(\ref{eq:Gammadef}). However, before we
embark on a detailed study of how these predictions depend on some
specific aspects of the halo occupation model used, it is important to
stress that the results in Fig.~3 are in real-space. Unfortunately,
because of redshift-space distortions and projection effects,
real-space correlation functions are extremely difficult, if not
impossible, to obtain observationally.  We therefore first recast our
bias functions into two-dimensional, projected analogues, which are
more easily accessible observationally. We start by defining the
matter-matter, galaxy-matter, and galaxy-galaxy projected surface
densities as
\begin{equation}
\Sigma_{\rm xy}(r_\rmp) = 2  {\bar \rho} \, \int_{r_\rmp}^{\infty} 
\left[1+\xi_{\rm xy}(r)\right] \, { r \rmd r \over \sqrt{r^2 - r_\rmp^2}}\,,
\end{equation}
where `x' and `y' stand either for `g' or `m', and $r_{\rm p}$ is the
\emph{projected} separation.  We also define $\overline{\Sigma}_{\rm
  xy}(<r_\rmp)$ as its average inside $r_{\rm p}$, i.e.
\begin{equation}\label{averageSigma}
\bar{\Sigma}_{\rm xy}(<r_\rmp)  = \frac{2}{r^2_\rmp}
\int_{0}^{r_\rmp}\Sigma_{\rm xy}(R') R' dR'\,,
\end{equation}
which we use to define the excess surface densities
\begin{equation}\label{ESDxy}
\Delta \Sigma_{\rm xy}(r_\rmp) = \bar{\Sigma}_{\rm xy}(<r_\rmp) - 
\Sigma_{\rm xy}(r_\rmp)\,.
\end{equation}
These can subsequently be used to define the projected, 2D analogues
of Eqs.~(\ref{eq:bgaldef})-(\ref{eq:Gammadef}) as
\begin{equation}\label{eq:bgaldef2}
b_\rmg(r_\rmp) \equiv \sqrt{\frac{\Delta \Sigma_{\rm gg}(r_\rmp)}
{\Delta \Sigma_{\rm mm}(r_{\rm p})}} \,, 
\end{equation}
\begin{equation}\label{eq:CCCdef2}
{\cal R}_{\rm gm}(r_\rmp) \equiv \frac{\Delta \Sigma_{\rm gm}(r_\rmp)}
{\sqrt{\Delta \Sigma_{\rm gg}(r_\rmp)\, \Delta \Sigma_{\rm mm}(r_\rmp)}}\,,
\end{equation}
and
\begin{equation}\label{eq:Gammadef2}
\Gamma_{\rm gm}(r_\rmp) \equiv \frac{b_\rmg(r_\rmp)}
{{\cal R}_{\rm gm}(r_\rmp)} =
\frac{\Delta \Sigma_{\rm gg}(r_\rmp)}{\Delta \Sigma_{\rm gm}(r_\rmp)}\,, 
\end{equation}
In what follows we shall refer to these as the `projected bias
functions'.

Note that in the case of the galaxy-dark matter cross correlation, the
excess surface density $\Delta \Sigma_{\rm gm}(r_\rmp) =
\gamma_\rmt(r_\rmp) \,\Sigma_{\rm crit}$, where $\gamma_\rmt(r_\rmp)$
is the tangential shear which can be measured observationally using
galaxy-galaxy lensing, and $\Sigma_{\rm crit}$ is a geometrical
parameter that depends on the comoving distances of the sources and
lenses. In the case of the galaxy-galaxy autocorrelation we can write
that
\begin{equation}
\Delta \Sigma_{\rm gg}(r_\rmp) = \bar{\rho} \left[
{2 \over r^2_\rmp} \int_0^{r_\rmp} w_\rmp(R') \, R' \, \rmd R' 
- w_\rmp(r_\rmp)\right]\,,
\end{equation}
from which it is immediately clear that $\Delta_{\rm gg}(r_\rmp)$ is
straightforwardly obtained from the projected correlation function
$w_\rmp(r_\rmp)$, which is routinely measured in large galaxy redshift
surveys. Finally, in the case of the matter-matter autocorrelation,
the quantity $\Delta \Sigma_{\rm mm}(r_{\rm p})$ can be obtained
observationally if accurate cosmic shear measurements are
available. Since the cosmic shear measurements are the most
challenging, the parameter $\Gamma_{\rm gm}(r_\rmp)$, which is
independent of $\Delta\Sigma_{\rm mm}(r_\rmp)$, is significantly
easier to determine observationally than either $b_\rmg(r_\rmp)$
and/or $\calR_{\rm gm}(r_\rmp)$. In fact, current clustering and
galaxy-galaxy lensing data from the SDSS is already of sufficient
quality to allow for reliable measurements of $\Gamma_{\rm
  gm}(r_\rmp)$ for different luminosity bins. The purpose of this
paper, however, is not to perform such measurements, but rather to
provide theoretical guidance on how to constrain different aspects of
galaxy biasing by exploiting the wealth of information encoded in the
scale dependence of the (projected) galaxy bias functions.

\section{Impact of halo occupation assumptions on galaxy biasing}
\label{sec:impact}
 
We now investigate how modifications of the halo occupation
statistics, modelled via the CLF, impact the projected bias functions
$b_\rmg(r)$, $\calR_{\rm gm}(r)$ and $\Gamma_{\rm gm}(r)$ defined in
Eqs.~(\ref{eq:bgaldef2})-(\ref{eq:Gammadef2}). We first study
modifications regarding the way central galaxies occupy dark matter
haloes, followed by an indepth study of the impact of various aspects
of satellite occupation statistics.
\begin{figure*}
\psfig{figure=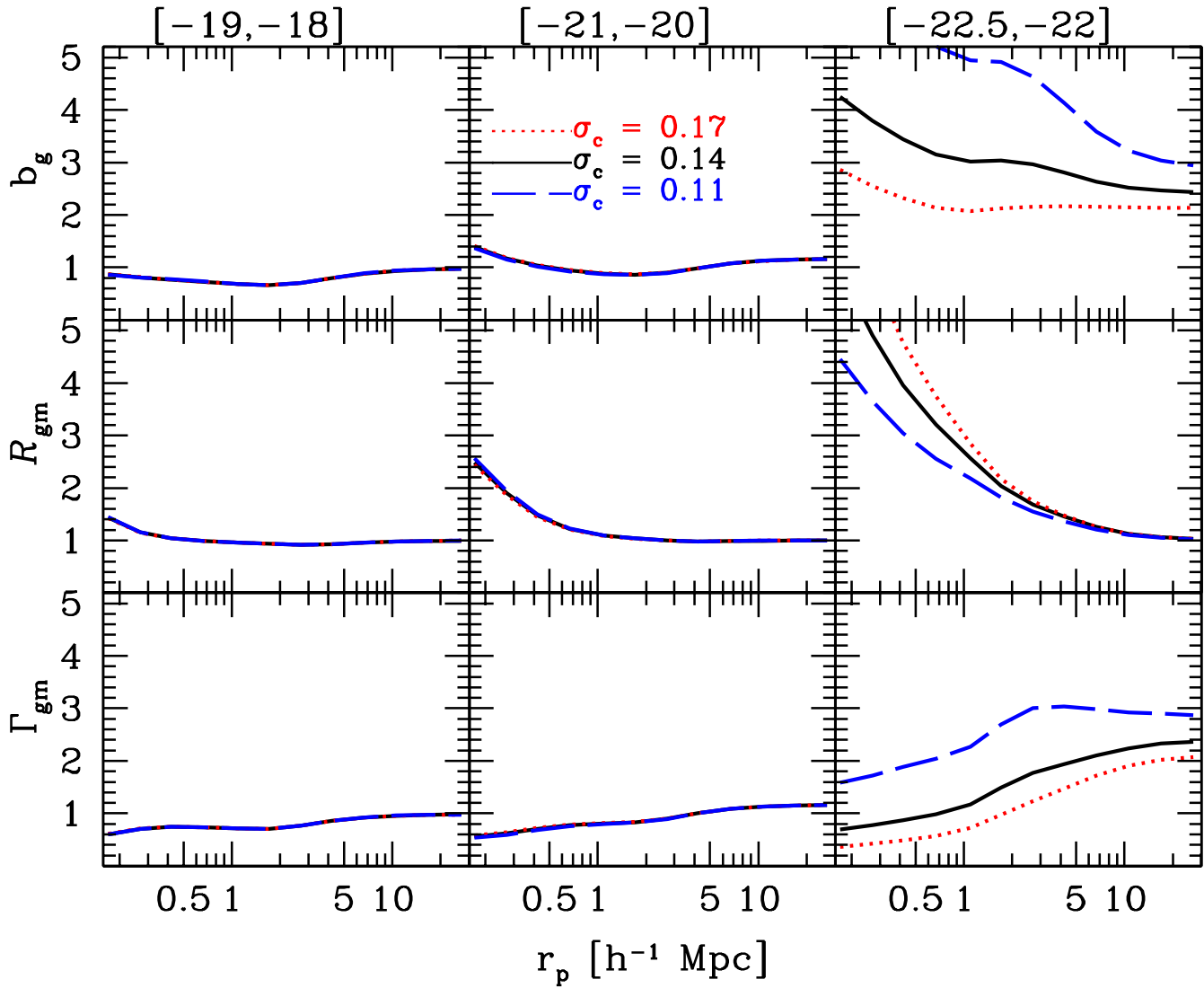,width=\hdsize}
\caption{
  Same as in Fig.~\ref{fig:fig4}. 
  The reference model is
  indicated with the black solid lines, whereas other lines refer to
  models with larger ($\sigma_{\rm c}=0.17$, red dotted lines) or smaller
  ($\sigma_{\rm c}=0.11$, blue dashed lines) scatter in the luminosity-halo
  mass relation (see eq.~[\ref{phi_c}] and discussion in \S5.2).  }
\label{fig:sigma}
\end{figure*}

\subsection{The Relative Contribution of Centrals}
\label{sec:relcont}

We start by exploring the role of central galaxies by comparing models
in which the relative contribution of centrals is progressively
suppressed via the parameter $\eta$ (see \S\ref{sec:example}).  Figure
4 shows the scale dependence of the projected bias functions $b_{\rm
  g}, {\cal R}_{\rm gm}, \Gamma_{\rm gm}$ for the same luminosity bins
as in Figure 3. The bias functions are plotted as a function of the
projected radius, $r_{\rm p}$, covering the range [0.1, 30]$ h^{-1}$
Mpc. Results are shown for the reference model ($\eta=1$, black solid
lines) and for models with an increasingly lower contribution from
centrals ($\eta=2, 5, 10, 100$, as indicated).  Overall, the trends
are very similar to those seen in Fig.~3. For instance, the reference
model once again shows that the magnitude of scale dependence
increases with luminosity. However, upon closer examination some
important differences become apparent, which arise from projecting and
integrating the two-point correlation functions, which are the
operations performed in order to compute the excess surface densities
given by Eq.~[\ref{ESDxy}]. An important difference is that the galaxy
bias, $b_\rmg$, remains scale dependent out to much larger radii; in
the highest luminosity bin, significant scale dependence remains
visible out to $\sim 20 h^{-1}\Mpc$. This is very different from the
case of $b_\rmg^{\rm 3D}$ which becomes scale-independent at $\sim 5
h^{-1}\Mpc$, independent of the value of $\eta$.  This difference is a
consequence of the integration~(\ref{averageSigma}), which mixes in
signal from small scales. This mixing also smears out the sharp
features in the 1-halo to 2-halo transition region that are present in
their 3D analogues. Hence, although the projected bias functions have
the advantage that they are observationally accessible, their
interpretation is less straightforward. Nevertheless, as we will see
below, different aspects of the halo occupation statistics impact the
projected bias functions in sufficiently different ways that they
still provide valuable insight into the nature of galaxy biasing.

\subsection{Scatter in the Luminosity-Halo Mass Relation}

Throughout this paper, the number of central galaxies with luminosity
$L$ that reside in a halo of mass $M$ is modelled as a log-normal
distribution (see Eq.~[\ref{phi_c}]), whose standard deviation,
$\sigma_{\rm c}$, indicates the scatter in luminosity at a given halo
mass. Following \citet{2009MNRAS.394..929C}, and motivated by the
results of \citet{2009MNRAS.392..801M} based on satellite kinematics,
we assume that $\sigma_{\rm c}$ is independent of halo
mass. \citet{2009MNRAS.394..929C} obtained $\sigma_{\rm c}=0.14$,
which is the value we adopt in our fiducial reference model.

Figure~\ref{fig:sigma} shows the projected galaxy bias functions for
the reference model ($\sigma_{\rm c}=0.14$, black solid lines) and for
two models with $\sigma_{\rm c}=0.17$ (red dotted lines) and
$\sigma_{\rm c}=0.11$ (blue dashed lines), respectively. All other
parameters are kept unchanged. Only the brightest bin reveals
appreciable changes in the bias functions. This is most easily
understood by examining the upper panels of Fig.~\ref{fig:HODs}, which
show how changes in $\sigma_\rmc$ impact the occupation statistics.
For the two fainter bins, changes in $\sigma_\rmc$ only have a very
mild impact on the HODs. This in turn is a consequence of the fact
that fainter galaxies reside, on average, in less massive haloes, and
therefore probe the low mass end of the halo mass function. In this
regime, the halo mass function is a power law and variations in the
scatter in $\Phi_{\rm c}(L|M)$ have little effect on the resulting
mass of the {\it average} halo hosting these galaxies. Conversely,
brighter galaxies probe the high mass end of the halo mass function,
which reveals an exponential decline.  Thus, variations in the scatter
in $\Phi_{\rm c}(L|M)$ strongly affect the resulting mass of the {\it
  average} halo hosting these galaxies (see upper right-hand panel of
Fig.~\ref{fig:HODs}), which reflects itself in a change of $b_\rmg$.
In particular, increasing (decreasing) $\sigma_\rmc$ strongly
suppresses (boosts) the bias $b_\rmg(r)$.

An interesting aspect of scatter (i.e., stochasticity) in the
occupation statistics of central galaxies, is that it impacts the
2-halo term, i.e., changes in $\sigma_\rmc$ have an impact on $b_\rmg$
and $\Gamma_{\rm gm}$, but not $\calR_{\rm gm}$, on large scales. This
is a consequence of the fact that changes in $\sigma_\rmc$ impact
$\langle N_\rmc|M\rangle$ (see discussion in \S\ref{sec:censat}). Note
that this is not the case for the stochasticity of satellite galaxies,
which only impacts the bias functions on small scales (where the
1-halo term dominates).

\subsection{The First Moment of $P(N_\rms|M)$}

The first moment, $\langle N_\rms|M \rangle$, of $P(N_\rms|M)$, for
any luminosity bin $[L_1,L_2]$, is completely specified by the CLF
(see \S~\ref{sec:example}). As shown in Fig.~\ref{fig:HODs}, the slope
of $\langle N_\rms|M \rangle$ is very sensitive to changes in the
parameter $\alpha_\rms$, which controls the faint-end slope of
$\Phi_\rms(L|M)$: in general, smaller (i.e., more negative) values of
$\alpha_\rms$ result in steeper $\langle N_\rms |M \rangle$. As such,
$\alpha_\rms$ is a parameter that controls the non-linearity of the
HOD. Data from clusters and galaxy groups typically indicate values
for $\alpha_\rms$ in the range $-1.5 \lta \alpha_\rms \lta -0.9$
(e.g., \citealt{2002MNRAS.329..385B, 2002MNRAS.335..712T,
  2004MNRAS.355..769E, YMB08})

\begin{figure*}
\psfig{figure=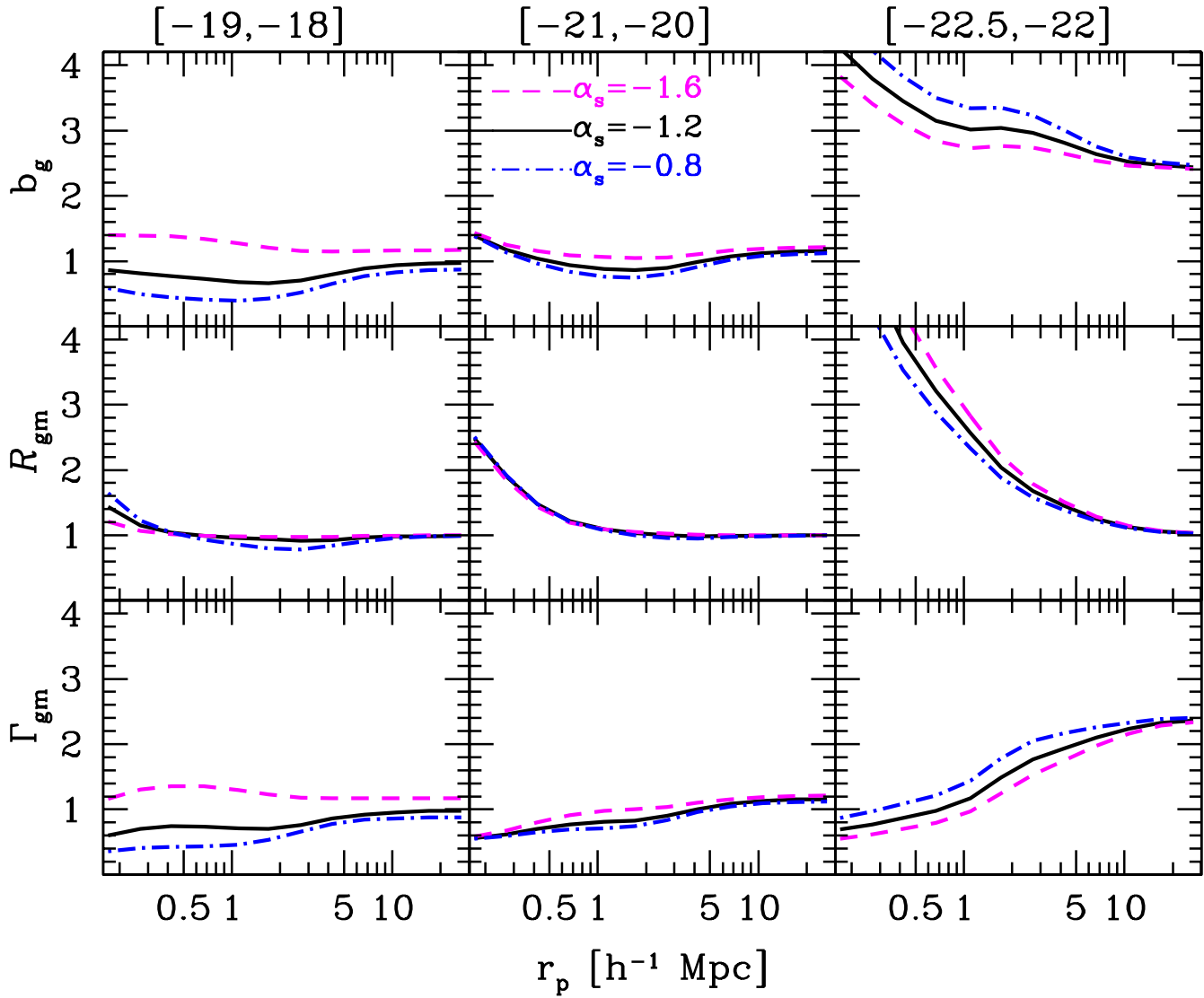,width=\hdsize}
\caption{
  Same as in Fig.~\ref{fig:fig4}.
  The reference model is
  indicated with the black solid lines, whereas other lines refer to
  models with lower ($\alpha_{\rm s}=-1.6$, magenta dashed lines) or higher
  ($\alpha_{\rm s}=-0.8$, blue dot-dashed lines) value of the low mass end
  power-law index of the CLF (see eq.[30] and discussion in \S5.3).  }
\label{fig:alpha}
\end{figure*}

Figure~\ref{fig:alpha} shows the projected bias functions for the
reference model (which has $\alpha_\rms = -1.2$) as well as for two
variations with $\alpha_\rms = -0.8$ and $-1.6$, as indicated. All
other model parameters are kept unchanged.  Note how smaller values of
$\alpha_\rms$ increase (decrease) the bias parameter $b_\rmg$ for
faint (bright) galaxies.  For faint galaxies, this is easy to
understand; a more negative $\alpha_\rms$ results in a steeper
$\langle N_\rms|M \rangle$, which implies that satellites, on average,
reside in more massive (i.e., more strongly biased) haloes. Since the
satellite fraction decreases with increasing luminosity, the impact of
changes in $\alpha_\rms$ become smaller for brighter galaxies.

\subsection{The Second Moment of $P(N_\rms|M)$}

The CLF does not specify the second moment, $\langle N_\rms^2|M
\rangle$, of the satellite occupation distribution. Rather, it is
often assumed that $P(N_\rms|M)$ follows a Poisson distribution
\begin{equation}
P(N_\rms|M) = \frac{\lambda^{N_\rms} \exp{[-\lambda]}}{N_\rms!} \, ,
\end{equation} 
where $\lambda = \langle N_\rms|M\rangle$ is the first moment of the
distribution. Recall that for a Poisson distribution all moments can
be derived from the first moment. In particular, $\langle
N_\rms(N_\rms-1)|M\rangle = \langle N_\rms|M\rangle^2$.  The
assumption that $P(N_\rms|M)$ is (close to) Poissonian has strong
support from galaxy group catalogs (e.g., \citealt{YMB08}) and from
numerical simulations, which show that dark matter subhaloes (which
are believed to host satellite galaxies) also follow Poisson
statistics (\citealt{2004ApJ...609...35K}). However, a number of
studies have shown that there may be small but significant deviations
from pure Poisson statisitcs (e.g., \citealt{2004MNRAS.355.1010P,
  2005MNRAS.359.1029V, 2010MNRAS.404..502G, 2010MNRAS.406..896B,
  2011ApJ...743..117B}). Hence, we investigate how deviations from
Poisson, as parameterized via the parameter $\beta$ (see eq. 25),
impact on the (projected) bias functions.
\begin{figure*}
\psfig{figure=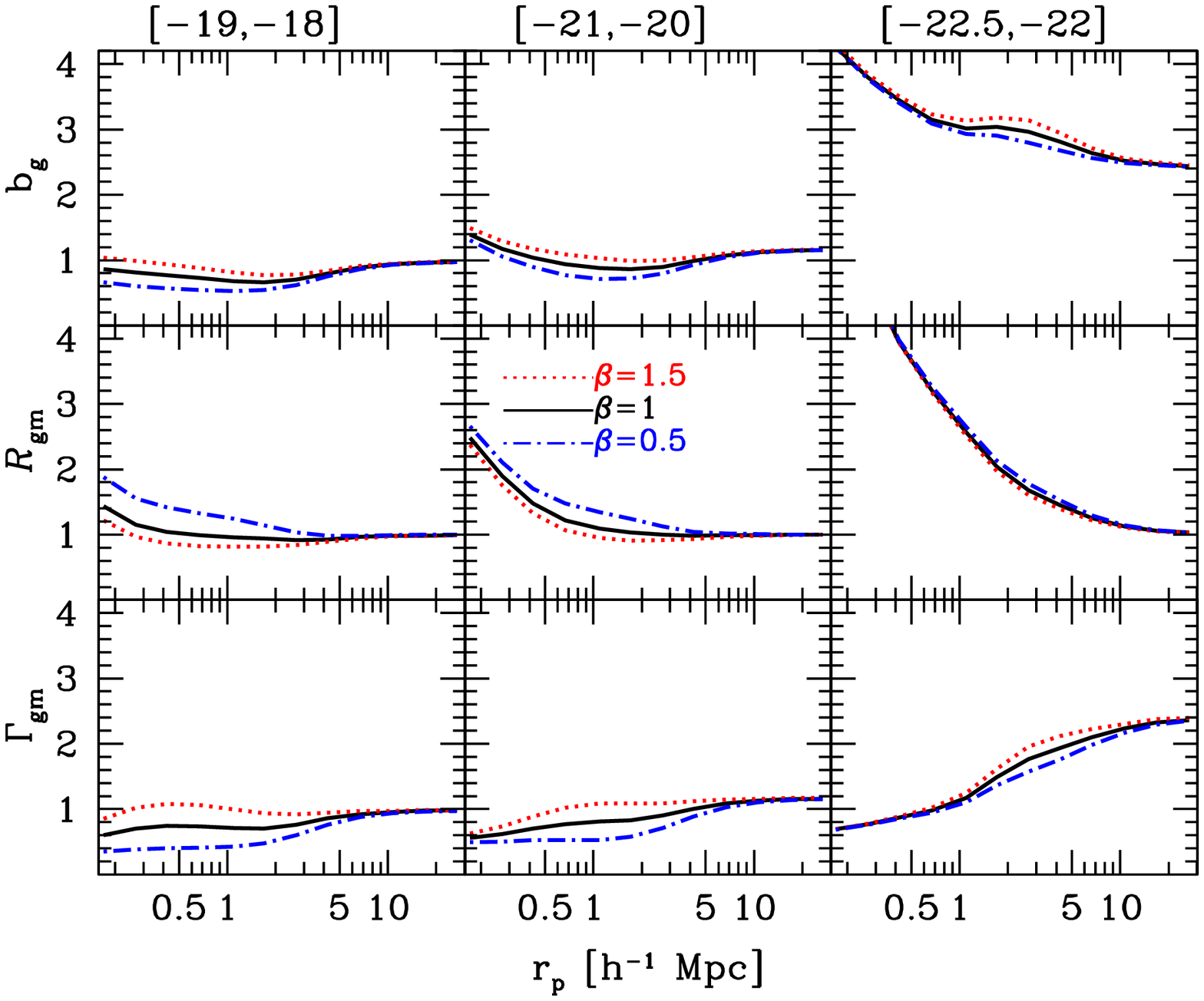,width=\hdsize}
\caption{
  Same as in Fig.~\ref{fig:fig4}. 
  The reference model is
  indicated with the black solid lines, whereas other lines refer to
  models with larger ($\beta=1.5$, red lines) or smaller ($\beta=0.5$,
  blue lines) Poisson parameter (see eq.[25] and discussion in \S5.4).  }
\label{fig:Poisson}
\end{figure*}

Figure~\ref{fig:Poisson} shows the projected bias functions for the
reference model ($\beta =1$, corresponding to $P(N_\rms|M)$ being
Poissonian) as well as for two variations with $\beta = 0.5$ and
$1.5$, as indicated. All other model parameters are kept
unchanged. The parameter $\beta$ only affects the 1-halo
satellite-satellite term of the galaxy-galaxy correlation function.
This term typically has a maximum contribution close to the 1-halo to
2-halo transition region, which is therefore the radial interval that
is most sensitive to changes in $\beta$.  Since brighter galaxies
reside in more massive (and therefore more extended) haloes, changes
in $\beta$ impact the (projected) bias functions on larger scales for
brighter galaxies. Also, since the satellite fraction increases with
decreasing luminosity, the impact of changes in $\beta$ is larger for
less luminous galaxies. Overall, though, changes in $\beta$ of 50
percent only have a fairly modest impact on the (projected) bias
functions. Since $\beta$ is unlikely to differ from unity by more than
$\sim 20$ percent, we conclude that potential deviations from Poisson
statistics are unlikely to have a significant effect on galaxy
biasing.

\subsection{The Spatial Distribution of Satellites}

\begin{figure*}
\psfig{figure=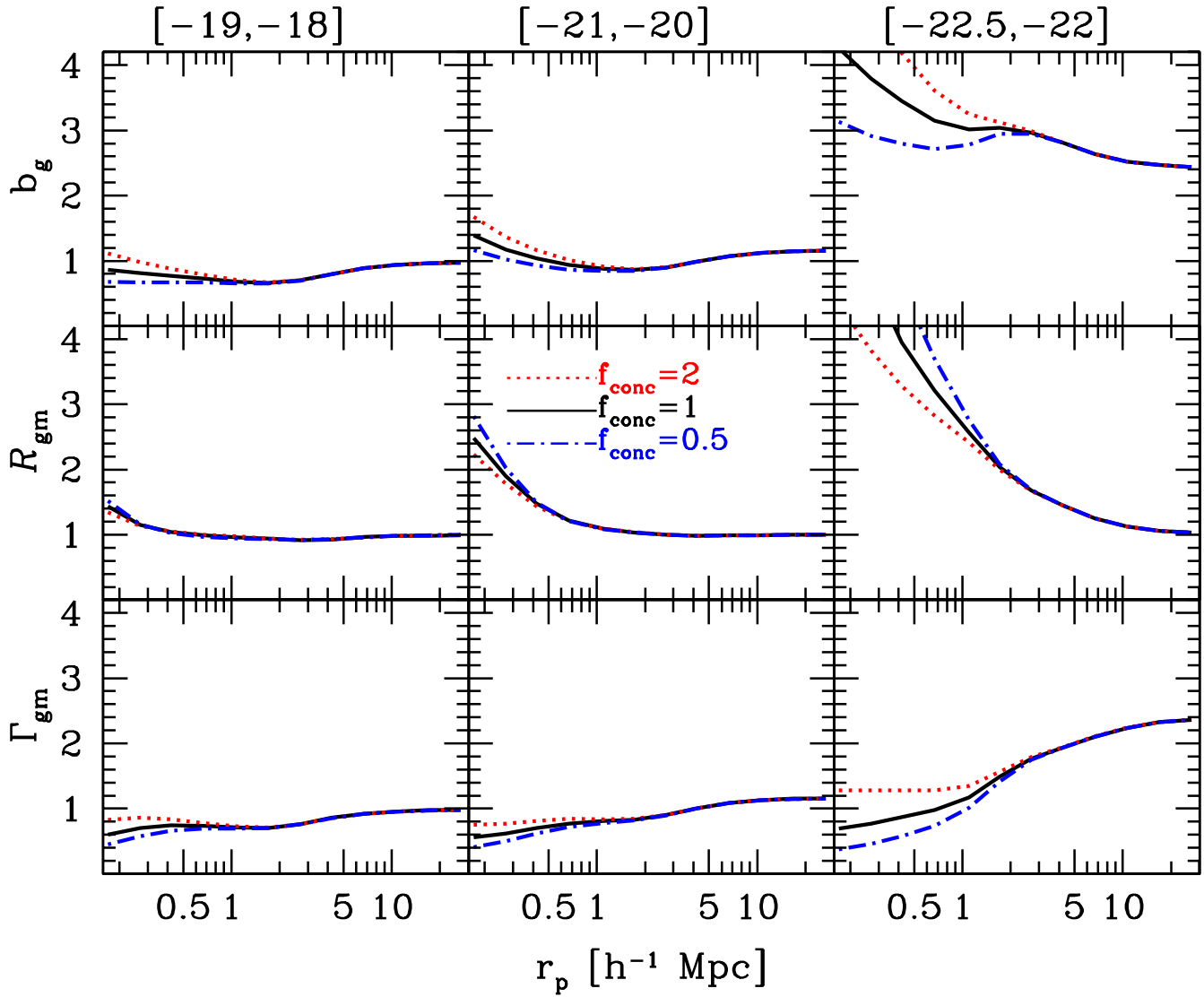,width=\hdsize}
\caption{
  Same as in Fig.~\ref{fig:fig4}. 
  The reference model is
  indicated with the black solid lines, whereas other lines refer to
  models with higher ($f_{\rm conc}=2$, red dotted lines) or lower ($f_{\rm
    conc}=0.5$, blue dot-dashed lines) value of the ratio between dark matter and
  galaxy concentration, $f_{\rm conc}=c_{\rm m}/c_{\rm g}$.  (see
  eq.~[\ref{eq:fconc}] and discussion in \S5.5).}
\label{fig:fconc}
\end{figure*}

In our fiducial reference model it is assumed that the number density
distribution of satellites within dark matter haloes is identical to
that of dark matter particles; i.e., we assume that
$\tilde{u}_\rms(k|M) = \tilde{u}_\rmh(k|M)$. The dark matter density
profiles are modelled as NFW (\citealt{1997ApJ...490..493N}) profiles,
with a concentration mass relation, $c_\rmh(M)$, given by
\citet{2007MNRAS.378...55M}. Whether the assumption that the number
density distribution of satellite galaxies is well described by the
same NFW profile, and with the same concentration-mass relation, is
still unclear. In particular, numerous studies have come up with
conflicting claims (e.g., \citealt{1986ApJ...300..557B,
  1997ApJ...478..462C, 2000AJ....119.2038V, 2004ApJ...610..745L,
  2005MNRAS.356.1233V, 2005MNRAS.362..711Y, 2009A&A...494..867C,
  2009MNRAS.392..917M, 2011ApJ...731...44N, 2011arXiv1108.1195W,
  2012arXiv1201.1296G}). We therefore examine the impact of changing
the number density profiles of satellites, which we parameterize via
\begin{equation}\label{eq:fconc}
f_{\rm conc} = c_\rms/c_\rmh\,,
\end{equation} 
where $c_\rms$ is the concentration parameter of the satellite number
density profile. Note that $f_{\rm conc}=1$ for our fiducial reference
model. Figure~\ref{fig:fconc} shows how changes in $f_{\rm conc}$
impact the projected bias functions. The various curves correspond to
$f_{\rm conc} = 0.5$, $1.0$ and $2.0$, as indicated. All other model
parameters are the same as for the reference model. As expected,
changing the radial number density profile of satellite galaxies only
affects the bias functions on small scales where the 1-halo term
dominates. In general, a more centrally concentrated distribution of
satellites (i.e., larger $f_{\rm conc}$) results in a larger galaxy
bias and smaller CCC on small scales.  Since the impact of $f_{\rm
  conc}$ is restricted to smaller scales than most other changes in
the halo occupation statistics, accurate measurements of the
(projected) bias functions on small scales holds excellent potential
for constraining the radial number density profiles of satellite
galaxies.

\section{Observational Perspective}
\label{sec:observations}

\subsection{Probing the Scale Dependence of Galaxy Bias}
\label{sec:gbsmsc}

Constraining the non-linearity and stochasticity of galaxy bias with
observations is a non-trivial task. The main reason is that it
requires measurements of the fluctuations in the matter density (or
$n$-point statistics thereof), which are difficult to obtain. In the
absence of such information, however, one can still make some
progress.  For example, one can test for linearity by measuring
higher-order statistics of the galaxy distribution, such as the
bi-spectrum (e.g., \citealt{1999ApJ...521L..83F, 2002ApJ...570...75S,
  2002MNRAS.335..432V}). This method, though, yields no information
regarding stochasticity in the relation between galaxies and
matter. Alternatively, \citet{1999ApJ...518L..69T} suggested a method
to constrain the non-linearity and/or stochasticity by comparing
different samples of galaxies (i.e., different luminosity bins,
early-types {\it vs.}  late-types, etc.)  Quoting directly from their
paper: ``If two different types of galaxies are both perfectly
correlated with the matter, they must also be perfectly correlated
with each other''. Hence, if one can establish that the correlation
between the two samples is imperfect, than galaxy bias has to be
non-linear and/or stochastic for at least one of the two
samples. Different implementations of this idea have been used by a
number of authors (e.g., \citealt{2000ApJ...544...63B,
  2005MNRAS.356..456C, 2005MNRAS.356..247W, 2007ApJ...664..608W,
  2008MNRAS.385.1635S, 2011ApJ...736...59Z}), with the general result
that bias cannot be linear and deterministic.

A third method to constrain the non-linearity of galaxy bias without
direct measurements of the matter fluctuations was proposed by
\citet{2000ApJ...540...62S}. Here the idea is to measure the
probability distribution function (PDF) $\calP(\delta_\rmg)$ of the
galaxy field using counts-in-cells measurements, and to compare that
with (log-normal) {\it models} for the PDF $\calP(\delta_\rmm)$ of
mass fluctuations. Under the assumption that the bias relation between
galaxies and matter is deterministic and monotonic, one can use these
two PDFs to infer the conditional mean bias relation $\langle
\delta_\rmg | \delta_\rmm \rangle$. This method has been applied to
the first epoch VIMOS VLT deep survey (VVDS;
\citealt{2005A&A...439..845L}) by \citet{2005A&A...442..801M} and to
the zCOSMOS survey (\citealt{2007ApJS..172...70L}) by
\citet{2011ApJ...731..102K}. Both studies find clear indications that
bias is non-linear. However, this method has a number of important
shortcomings: it assumes that bias is deterministic, is
cosmology-dependent, and, because it has to filter the galaxy
distribution (typically on scales of $5 - 10 h^{-1} \Mpc$) is carries
no information of galaxy bias on small scales. As we have shown, these
are the scales that carry most information regarding the non-linearity
and stochasticity of galaxy bias.

Arguably the most promosing method to measure non-linearity and
stochasticity in galaxy bias, which does not require any assumptions
regarding cosmology or bias, is to use a combination of galaxy
clustering and gravitational lensing. The latter is the only direct
probe of the matter distribution in the Universe, and allows for
direct measurements of the galaxy-matter cross correlation (via
galaxy-galaxy lensing) as well as the matter auto-correlation (via
cosmic shear). When combined with the galaxy-galaxy correlation, these
measurements yield the bias functions $b_\rmg(r)$, $\calR_{\rm gm}(r)$
and $\Gamma_{\rm gm}(r)$ discussed in this paper. As elucidated above,
the scale dependence of these bias functions holds great promise to
gain valuable insight into the origin of non-linearity and
stochasticity of galaxy bias.

\citet{2003MNRAS.346..994P} applied these ideas to the VIRMOS-DESCART
cosmic shear survey (\citealt{2000A&A...358...30V}). Using
deprojection techniques, they computed the 3D power spectra for dark
matter and galaxies, as well as their cross correlation. They find a
weak indication of scale dependence with $\calR^{\rm 3D}_{\rm gm}$
decreasing with scale, in qualitative agreement with the predictions
shown in \S\ref{sec:impact}. \citet{1998A&A...334....1V} and
\citet{1998ApJ...498...43S} proposed a somewhat alternative
application of this method based on aperture masses and aperture
number counts. The aperture mass, $M_{\rm ap}(\theta)$, is defined as
\begin{equation}
M_{\rm ap}(\theta) = \int {\rm d}^2 \phi \, U(\phi) \, 
\delta^{\rm 2D}_{\rm m}(\phi) \, ,
\end{equation}
(\citealt{1998MNRAS.296..873S}), where $\delta^{\rm 2D}_{\rm m}$ is
the projected mass overdensity and $U(\phi)$ is a compensated filter,
i.e. $\int {\rm d}\phi \, \phi \, U(\phi) = 0$ and $U(\phi) = 0$ for
$\phi > \theta$. Similarly, for galaxies one defines the aperture
number count
\begin{equation}
N_{\rm ap}(\theta) = \int {\rm d}^2 \phi \, U(\phi) \,
\delta^{\rm 2D}_{\rm g}(\phi) \, ,
\end{equation}
with $ \delta^{\rm 2D}_{\rm g}$ the projected galaxy overdensity.  The
mass autocorrelation function $\langle M^2_{\rm ap}(\theta) \rangle$
can be derived from the observed ellipticity correlation functions,
the angular two-point correlation function of galaxies yields $\langle
N^2_{\rm ap}(\theta) \rangle$, and the ensemble-averaged tangential
shear as a function of radius around a sample of lenses (galaxy-galaxy
lensing signal) can be used to derive $\langle N_{\rm ap}(\theta)
M_{\rm ap}(\theta) \rangle$ (\citealt{2002A&A...393..369V,
  2002ApJ...577..604H}). Relating these variances at a given smoothing
aperture, $\theta$, one obtains the aperture bias functions
\begin{equation}\label{eq:bap}
b_{\rm ap}(\theta) \propto \frac{\langle N^2_{\rm ap}(\theta) \rangle}
{\langle M^2_{\rm ap}(\theta) \rangle} \,,
\end{equation}
and
\begin{equation}\label{eq:Rap}
\calR_{\rm ap}(\theta) \propto 
\frac{\langle N_{\rm ap}(\theta) M_{\rm ap}(\theta) \rangle}
{\sqrt{\langle N^2_{\rm ap}(\theta) \rangle 
\langle M^2_{\rm ap}(\theta) \rangle}}\,,
\end{equation} 
where the constants of proportionality depend in principle on the
distribution of galaxies and on the assumed cosmological model
(\citealt{1998A&A...334....1V}), although minimally within the current
uncertainties on cosmological parameters
(\citealt{2011ApJS..192...18K}). 

\begin{figure}
\psfig{figure=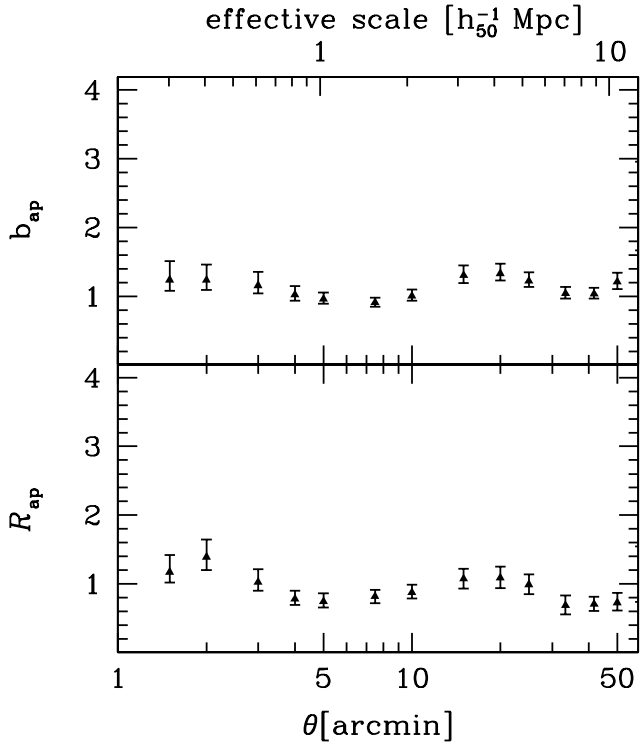,width=\hssize}
\caption{ The scale dependence of the galaxy bias (top panel) and the
  cross-correlation coefficient (bottom panel) as measured with the aperture mass rechnique
  in the RCS survey (see discussion in \S6.1),
  properly accounting for PSF anisotropy (following
  \citealt{2005A&A...429...75V}).  }
\label{fig:Hoekstra}
\end{figure}

The aperture-based method has been applied to data from the RCS
(\citealp{2002ASPC..257..109Y}) in \citet{2001ApJ...558L..11H} and in
combination with data from the VIRMOS-DESCART survey in
\citet{2002ApJ...577..604H}.  Their results indicate that $b_{\rm ap}$
and $\calR_{\rm ap}$ are scale-dependent over scales $\sim 0.1 - 5
h^{-1}\Mpc$, but that the ratio $\Gamma_{\rm ap} \equiv b_{\rm
  ap}/\calR_{\rm ap}$ is nearly constant at $\Gamma_{\rm ap} \sim 1.1$
over this range. On scales of $\sim 0.5 h^{-1}\Mpc$, they find that
$b_{\rm ap} = 0.71^{+0.06}_{-0.05}$ and $\calR_{\rm ap} \sim
0.59^{+0.08}_{-0.07}$ (68\% confidence, assuming a flat $\Lambda$CDM
cosmology with $\Omega_\rmm=0.3$).  In Fig.~\ref{fig:Hoekstra}, we
show a revised version of the analysis published in
\citet{2002ApJ...577..604H}, which was based on the cosmic shear
analysis of \citet{2002A&A...393..369V}. However, as discussed in
\citet{2005A&A...429...75V}, these data suffered from PSF anisotropy
that was not corrected for. Figure~\ref{fig:Hoekstra} shows the new
results obtained using the same cosmology as in \citet{2002ApJ...577..604H},
the unchanged RCS measurements and the
\citet{2005A&A...429...75V} cosmic shear results\footnote{ Note that
  sample variance for both measurements has been ignored, which
  implies that the error budget is underestimated on large scales
  (above 10 arcmin).}. Compared to the results in
\citet{2002ApJ...577..604H}, the scale dependence on small scales has
been somewhat reduced. On scales of $\sim 0.5 h^{-1}\Mpc$, the new
analysis suggests $b_{\rm ap} = 0.96^{+0.09}_{-0.07}$ and $\calR_{\rm
  ap} = 0.74^{+0.12}_{-0.09}$ (68\% confidence). 

The finding that $b_{\rm ap}(\theta)$ and $\calR_{\rm ap}(\theta)$
have similar scale dependence so that $\Gamma_{\rm ap}(\theta)$ is
nearly scale independent (over the scales probed) seems at odds with
the predictions in \S\ref{sec:impact}. However, we caution that the
bias parameters based on aperture variances are different from the
ones investigated in this paper, which are based on excess surface
densities. Furthermore, lacking any redshift information, 
\citet{2002ApJ...577..604H} used a (apparent) magnitude selected sample 
which mixes galaxies of different 
intrinsic luminosity and different physical scales.
Hence, one cannot directly compare our model predictions
(Figs.~\ref{fig:fig4}-\ref{fig:fconc}) with the data in
Fig.~\ref{fig:Hoekstra}. 
On the other hand, the difference between
aperture variances and excess surface densities is mainly operational,
rather than conceptual, and it is difficult to imagine that they would
predict very different scale dependencies. In that respect it is
interesting that a more recent study by \citet{2012arXiv1202.6491J},
using the same aperture variance analysis, but applied to data from
the COSMOS survey (\citealt{2007ApJS..172....1S}), obtain $b_{\rm
  ap}(\theta)$ and $\calR_{\rm ap}(\theta)$ that more closely follow
the trends shown in Figs.~\ref{fig:fig4}-\ref{fig:fconc}. We intend to
interpret these aperture variance data within the context of halo
occupation statistics in a future publication.

Finally, we emphasize that the galaxy-matter cross correlation is a
first-order measure of the cosmic shear and therefore much easier to
measure than the matter auto-correlation, which is second-order (see
e.g., \citealt{1998ApJ...498...43S}).  Hence, the bias parameter
$\Gamma_{\rm gm}$ can typically be measured with significantly smaller
error bars than either $b_\rmg$ or $\calR_{\rm gm}$, simply because it
does not require the matter-matter correlation function. As we have
shown in this paper, the scale dependene of $\Gamma_{\rm gm}$, even the
one defined via the projected surface densities, contains a wealth of
information regarding the non-linearity and stochasticity of the halo
occupation statistics, and thus regarding galaxy formation.
\citet{2004AJ....127.2544S} used galaxy clustering and galaxy-galaxy
lensing data from the SDSS in order to measure $\Gamma^{\rm 3D}_{\rm
  gm}(r)$ for a large sample of $\sim 100,000$ galaxies. Using Abel
integrals, they deproject their data in order to estimate the 3D
galaxy-galaxy and galaxy-matter correlation functions. The resulting
$\Gamma^{\rm 3D}_{\rm gm}(r)$ is found to be roughly scale-invariant
over the radial range $0.2 h^{-1}\Mpc \lta r \lta 6.7 h^{-1} \Mpc$ at
a value of $(1.3 \pm 0.2) \, (\Omega_\rmm/0.27)$. Note that the
galaxies used in this measurement cover a wide range in magnitudes,
making it difficult to compare directly to the `predictions' in
Figs.~\ref{fig:fig4}-\ref{fig:fconc}.  It would be interesting to
repeat these measurements using narrower luminosity bins extracted
from the significantly larger and more accurate SDSS data samples
currently available.  We advocate to perform such an analysis using
the projected surface densities (Eq.~[\ref{ESDxy}]), rather than the
deprojected 3D quantities used by \citet{2004AJ....127.2544S}.

\subsection{Suppressing the Scale Dependence of Galaxy Bias}
\label{sec:Reyes}

For some applications, it is desirable to have bias functions with as
little scale-dependence as possible. As discussed in
\S\ref{sec:relcont}, because of the integration performed when
calculating the observable, projected bias functions, signal on small
scales is mixed-in on large scales. This causes the scale above which
the bias functions are scale-independent to increase. A
constraint on the amount of scale-dependence therefore means that a
large fraction of the data would have to be discarded. This can be
mitigated, however, by defining alternative bias functions that
circumvent mixing-in signal form small scales. For instance,
\citet{2010Natur.464..256R} used the quantities
\begin{equation}\label{Upsilon}
\Upsilon_{\rm xy}(r_\rmp) \equiv \Delta\Sigma_{\rm xy}(r_\rmp) - 
\left({r_{\rm min}\over r_\rmp}\right)^2 \Delta\Sigma_{\rm xy}(r_{\rm min})
\end{equation}
which are directly related to the excess surface densities defined in
Eq.~(\ref{ESDxy}), and where $r_{\rm min}$ is some fiducial length
scale. By rewriting Eq.~(\ref{Upsilon}) as
\begin{eqnarray}\label{Upsilon2}
\Upsilon_{\rm xy}(r_\rmp) & = & {2 \over r^2_\rmp} 
\int_{r_{\rm min}}^{r_\rmp} \Sigma_{\rm xy}(R') 
\,  R' \, \rmd R' \, - \Sigma_{\rm xy}(r_\rmp) \nonumber \\
& + & \left({r_{\rm min}\over r_\rmp}\right)^2 \Sigma_{\rm xy}(r_{\rm min})
\end{eqnarray}
it is immediately clear that $\Upsilon_{\rm xy}$ does not include any
contribution from length scales smaller than $r_{\rm min}$. Hence, one
could opt to define the projected bias
functions~(\ref{eq:bgaldef2})-(\ref{eq:Gammadef2}) using
$\Upsilon_{\rm xy}(r_\rmp)$ rather than $\Delta\Sigma_{\rm
  xy}(r_\rmp)$. In what follows we shall indicate these new bias
functions as $\widehat{b}_\rmg$, $\widehat{\calR}_{\rm gm}$ and
$\widehat{\Gamma}_{\rm gm}$, respectively.
\begin{figure*}
\psfig{figure=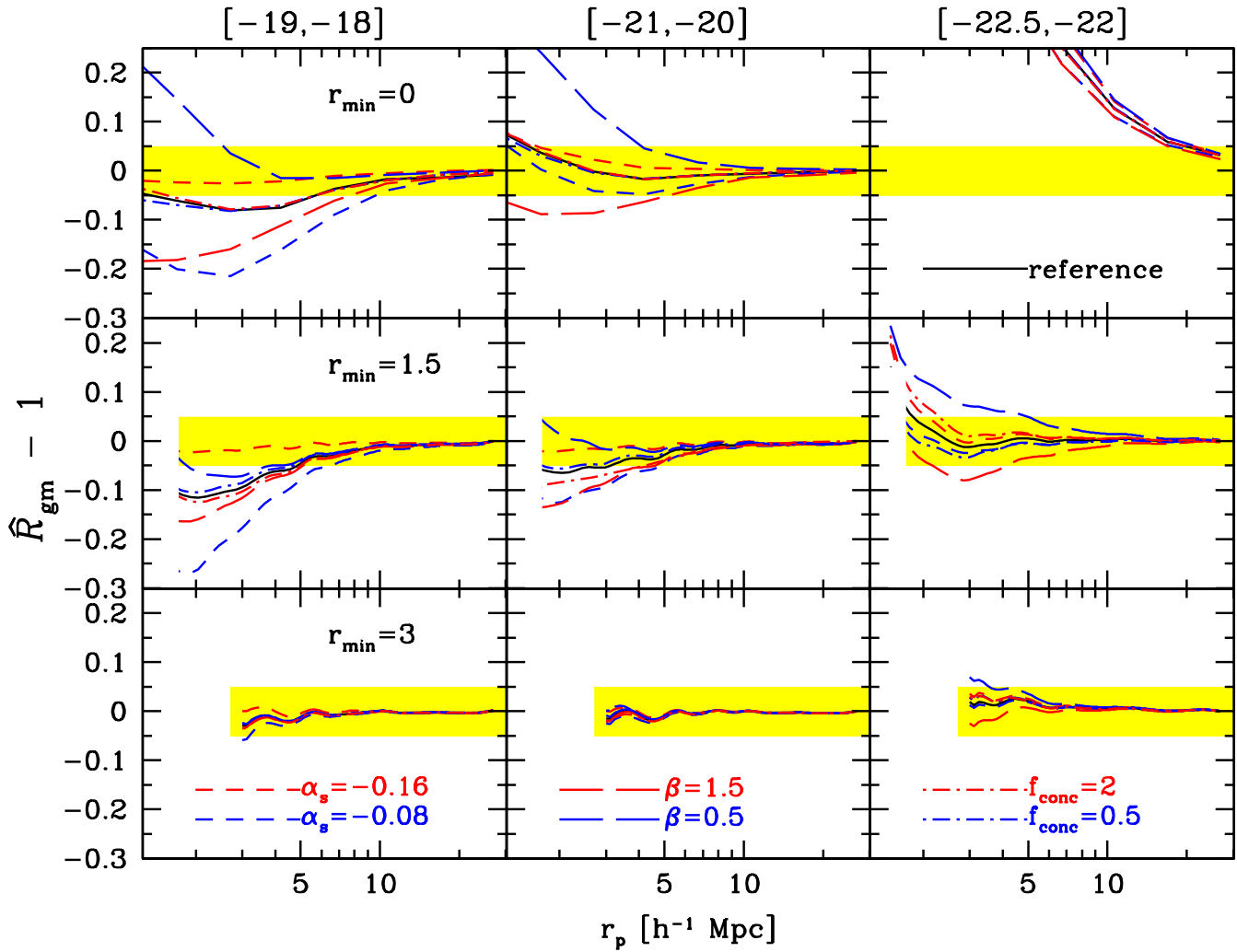,width=\hdsize}
\caption{The quantity $\widehat{\calR}_{\rm gm}(r_\rmp)
- 1$ for three different values of $r_{\rm min}$ (different rows), and
for three magnitude bins (different columns). The black solid curve
corresponds to our fiducial reference model, while other curves
correspond to several variations discussed
in \S\ref{sec:impact} (as indicated in the bottom panels).  The shaded
area indicates the region where scale dependence of 
$\widehat{\calR}_{\rm gm}(r_\rmp)$ yields values close to unity within 5\%.}
\label{fig:Reyes}
\end{figure*}

\citet{2010Natur.464..256R} used galaxy clustering, galaxy-galaxy
lensing and peculiar velocities of luminous red galaxies (LRGs) from
the SDSS to measure the probe of gravity
(\citealt{2007PhRvL..99n1302Z})
\begin{equation}
E_\rmG(r_\rmp) \equiv {1 \over \widehat{\beta}} 
{\Upsilon_{\rm gm}(r_\rmp) \over \Upsilon_{\rm gg}(r_\rmp)}
= f(\Omega_\rmm) \, \widehat{\calR}_{\rm gm}(r_\rmp)\,,
\end{equation}
where $\widehat{\beta} = f(\Omega_\rmm)/b_\rmg$ is the redshift
distortion parameter (not to be confused with the Poisson parameter
$\beta$), which can be measured from the redshift space correlation
function (e.g., \citealt{2006PhRvD..74l3507T}), $f(\Omega_\rmm)
\approx \Omega_\rmm^{0.55}$ is the logarithmic linear growth rate, and
$b_\rmg$ is the galaxy bias\footnote{Note that our definition of
  $E_\rmG$ differs from that of \citet{2010Natur.464..256R} by a
  factor $\Omega_{{\rm m},0}$, which is irrelevant for the discussion
  here.}. As long as $\widehat{\calR}_{\rm gm}(r_\rmp) = 1$, which can
be assured by picking a sufficiently large $r_{\rm min}$, it is clear
that $E_\rmG$ is a direct probe of $f(\Omega_\rmm)$, which is
sensitive to modifications of the law of gravity. In their analysis,
\citet{2010Natur.464..256R} adopt $r_{\rm min}=1.5 h^{-1}\Mpc$.

We now use our models to investigate the amount of scale dependence in
$\widehat{\calR}_{\rm gm}(r_\rmp)$ for various values of $r_{\rm
  min}$.  Fig.~\ref{fig:Reyes} plots $\widehat{\calR}_{\rm gm}(r_\rmp)
- 1$ for three different values of $r_{\rm min}$ (different rows), and
for three magnitude bins (different columns). The solid curve
corresponds to our fiducial reference model, while other curves
correspond to several variations with respect to this model discussed
in \S\ref{sec:impact} (as indicated in the bottom panels).  The shaded
area indicates the region where scale dependence of
$\widehat{\calR}_{\rm gm}(r_\rmp)$ affects the measurement of $E_\rmG$
at less than five percent.
 
In the upper panels we set $r_{\rm min}=0$, for which
$\widehat{\calR}_{\rm gm}(r_\rmp)$ reduces to $\calR_{\rm
gm}(r_\rmp)$. As we have already seen, this CCC reveals very
strong scale-dependence, especially for bright galaxies, making
$\calR_{\rm gm}(r_\rmp)$ useless for measuring $E_\rmG$.  For
$r_{\rm min} = 1.5 h^{-1} \Mpc$ (panels in middle row), this scale
dependence is drastically reduced, in particular for the brightest
galaxies. Note, though, that depending on the exact values of
$\alpha_\rms$, $f_{\rm conc}$ and $\beta$ the CCC
$\widehat{\calR}_{\rm gm}(r_\rmp)$ may still differ from unity at
the 10 to 20 percent level on small scales ($r_\rmp \sim r_{\rm
min}$).  However, adopting $r_{\rm min} = 3 h^{-1} \Mpc$ (lower
panels) yields $\widehat{\Gamma}_{\rm gm}(r_\rmp) = 1$ to better
than 5 percent accuracy for all luminosity bins, and with very
little dependence on uncertainties regarding the halo occupation
statistics. Hence, we conclude that the method used by
\citet{2010Natur.464..256R} succesfully suppresses the
scale-dependence of galaxy bias. For $r_{\rm min} = 1.5 h^{-1}\Mpc$,
which is the value adopted by \citet{2010Natur.464..256R} in their
analysis of LRGs in the SDSS, our models suggest, though, that there
may be some residual scale dependence on small scales at the 10-20
percent level, depending on detailed aspects of the halo occupation
statisitics. We find that robustly suppressing scale dependence in
$\widehat{\calR}_{\rm gm}(r_\rmp)$ to better than five percent
requires $r_{\rm min} \gta 3 h^{-1}\Mpc$.

Finally, we emphasize that the bias functions
$\widehat{b}_\rmg(r_\rmp)$, $\widehat{\calR}_{\rm gm}(r_\rmp)$ and
$\widehat{\Gamma}_{\rm gm}(r_\rmp)$ should only be used if one is
interested in suppressing scale-dependence. If, on the other hand, one
is interested in using two-point statistics to unveil the nature of
galaxy bias, which is the main goal of this paper, one should use the
projected bias functions (\ref{eq:bgaldef2})-(\ref{eq:Gammadef2})
instead.

\section{Summary  and Conclusions}
\label{sec:conclusions}

Studying galaxy bias is important for furthering our
understanding of galaxy formation, and for being able to use the
distribution of galaxies to constrain cosmology. Despite a large
number of (both observational and theoretical) studies, 
we still lack a useful framework for translating constraints on galaxy biasing
as inferred from observations into constraints on 
the theory of galaxy formation.

In an attempt to improve this situation, we have reformulated the
parameterization of non-linear and stochastic biasing introduced by
Dekel \& Lahav (1999; DL99) in the framework of halo occupation
statistics. The bias parameters introduced by DL99 relate the {\it
  smoothed} density field, $\delta_\rmg$, to the {\it smoothed} matter
field, $\delta_\rmm$.  This smoothing, however, has a number of
shortcomings. First, there is considerable loss of information 
on small scales (below the filtering scale), which, as we have shown 
in this paper, carry most information regarding galaxy bias. 
Second, the smoothing procedure is extremely sensitivity to (arbitrary) 
filtering scale: if the filtering scale is too large, the data has too little dynamic 
range to probe the non-linearity in the galaxy bias. A filtering scale that is too
small, on the other hand, results in too much shot-noise. Finally, 
the smoothing technique hampers  an intuitive link to the theory of galaxy formation.

Since galaxies are believed to form and reside in dark matter haloes,
it is far more natural, and intuitive, to define halo averaged
quantities (i.e., to adopt the host halo as the `filtering' scale).
This automatically suggest a reformulation of the DL99
parameterization in terms of halo occupation statistics. Since dark
matter haloes are biased entities themselves, such a formulation has
the additional advantage that the overall bias of galaxies has a
natural `split' in two components: (i) how haloes are biased with
respect to the dark matter mass distribution, and (ii) how 
galaxies are biased with respect to the haloes in which they reside. The
former has been addressed in detail with $N$-body simulations (e.g.,
\citealt{1996MNRAS.282..347M, 1998MNRAS.297..692C,
  1999ApJ...513L..99P, 1999MNRAS.304..767S}). The latter has been
the focus of this paper.

We have reformulated DL99 by replacing the filtered matter
overdensity, $\delta_\rmm$, with halo mass, $M$, and the filtered
galaxy overdensity, $\delta_\rmg$, with the occupation number, $N$.
Within this modified framework, the sources of non-linearity and
stochasticty in galaxy bias have logical and intuitive connections to
various aspect of galaxy formation. In particular, non-linearity
refers to deviations from $\langle N|M \rangle \propto M$, while
stochasticity refers to scatter in the probability distribution
function $P(N|M)$.  Based on our basic understanding of galaxy
formation, it is essentially impossible to have linear galaxy 
biasing. First of all, since galaxies of a given 
luminosity (or stellar mass) are
only expected to form in haloes above some minimum mass, one always
expects that $\langle N|M \rangle$ has some cut-off at low
$M$. Furthermore, there is no convincing reason why $\langle N|M
\rangle$ should scale linearly with halo mass above this mass
scale. In particular, since $\langle N|M \rangle = \langle N_\rmc|M
\rangle + \langle N_\rmc|M \rangle$, and $0 \leq \langle N_\rmc|M
\rangle \leq 1$, the contribution due to centrals typically gives rise
to strong non-linearity.  As for stochasticity, centrals contribute
from the fact that there is non-zero scatter in the relation between
halo mass and the luminosity of central galaxies (e.g.,
\citealt{2009MNRAS.392..801M}).  The contribution from satellites
comes from scatter in the halo occupation distribution
$P(N_\rms|M)$. Finally, an additional source of stochasticity may come
about if $N_\rmc$ and $N_\rms$ are not independent random variables.

A powerful method to probe the non-linearity and stochasticity of
galaxy bias is via two-point correlation functions. In particular,
non-linearity and stochasticty manifest themselves as scale dependence
in the bias parameter $b^{\rm 3D}_\rmg(r) \equiv \xi_{\rm
  gg}(r)/\xi_{\rm mm}(r)$, and the galaxy-matter cross correlation
parameter $\calR^{\rm 3D}_{\rm gm}(r) \equiv \xi_{\rm
  gm}(r)/\sqrt{\xi_{\rm gg}(r)\,\xi_{\rm mm}(r)}$. In this paper we
have proposed a modified set of bias parameters, $b_\rmg(r_\rmp)$ and
$\calR_{\rm gm}(r_\rmp)$, that are related to excess surface
densities, rather than real-space correlation functions. These have
the advantage that they can be inferred from data in a straightforward
manner, without the need for complicated, and noise-enhancing,
deprojection methods. In particular, combining galaxy clustering and
galaxy-galaxy lensing, it is straightforward to measure the ratio
$\Gamma_{\rm gm}(r_\rmp) \equiv b_\rmg(r_\rmp)/\calR_{\rm
  gm}(r_\rmp)$.

Using the halo model and a realistic halo occupation model, based on
the conditional luminosity function of \citet{2009MNRAS.394..929C}, we
have investigated how deviations from the linear and deterministic
biasing scheme manifest themselves in the scale dependence of
$b_\rmg$, $\calR_{\rm gm}$, and $\Gamma_{\rm gm}$. In particular, we
have compared predictions covering the spatial scale $0.1 \leq r_\rmp
\leq 30$ Mpc and for galaxies in three $r$-band magnitude bins;
[-19,-18], [-21,-20], and [-22.5,-22].  These choices aim at
bracketing the range of interest for current and forthcoming galaxy
surveys.  

We have shown that galaxy biasing is scale independent, with
$\calR^{\rm 3D}_{\rm gm}=1$, on large scales down to about $r \sim 2-5
h^{-1}\Mpc$.  The exact radius at which $\calR^{\rm 3D}_{\rm gm}$ (and
$b^{\rm 3D}_\rmg$) become scale dependent depends on luminosity, with
fainter galaxies remaining scale independent down to smaller scales.
This result is robust to all the model variations we have performed,
but it is only valid in real-space. When using the projected bias
parameters advocated here, which are more easily accessible
observationally, the associated quantity $\calR_{\rm gm}$ remains
scale dependent out to much larger radii (as far out as $\sim 20
h^{-1}\Mpc$). This comes about because the excess surface densities at
projected radius $r_\rmp$ contain information from all scales $r \leq
r_\rmp$. This `scale-mixing' can be avoided by using the {\it
  relative} excess surface densities 
  (see \citealt{2010PhRvD..81f3531B,2010Natur.464..256R}), 
which do not include any contribution
from length scales smaller than some fiducial radius $r_{\rm min}$.
Indeed, our models indicate that bias parameters based on these {\it
  relative} excess surface densities are scale independent at better
than 5 percent for $r_{\rm min} \geq 3 h^{-1}\Mpc$. For $r_{\rm min} =
1.5 h^{-1}\Mpc$, which is the value used by
\citet{2010Natur.464..256R} in their study to test the validity of GR,
we find residual scale dependence on small scales $r_\rmp \sim r_{\rm
  min}$ of the order of 20 percent\footnote{Since
  \citet{2010Natur.464..256R} did not use information on these small
  scales, their results are not influenced by this residual scale
  dependence.}.

On small scales ($r_\rmp \lta 2-5 h^{-1}\Mpc$), the bias functions
$b_\rmg(r_\rmp)$ and $\calR_{\rm gm}(r_\rmp)$ reveal strong scale
dependence. The detailed behavior of $b_\rmg(r_\rmp)$ and $\calR_{\rm
  gm}(r_\rmp)$ depends on (i) the luminosity of the galaxies in
question, with more luminous galaxies typically revealing stronger
scale-dependence, (ii) the detailed behavior of the occupation
statistics, $\langle N|M \rangle$, (iii) the scatter in the relation
between halo mass and the luminosity of central galaxies,
$\sigma_\rmc$, (iv) whether the satellite occupation distribution
$P(N_\rms|M)$ is Poissonian or not, and (v) the radial number density
distribution of galaxies within their host halo. For bright galaxies
($^{0.1}M_r - 5 \log h \lta -20$), the dominant effect giving rise to
the scale-dependence of $\calR_{\rm gm}(r_\rmp)$ is the precence of
central galaxies, which occupy very biased regions of their host
haloes. For fainter galaxies, $\calR_{\rm gm}(r_\rmp)$ is expected to
be close to unity, down to $r_\rmp \sim 0.2 h^{-1}\Mpc$, but with some
dependence on the Poisson parameter $\beta$. Finally, we stress that
$b_\rmg(r_\rmp)$ and $\calR_{\rm gm}(r_\rmp)$ for the brightest
galaxies are extremely sensitive to the amount of scatter,
$\sigma_\rmc$, in the relation between halo mass and the luminosity of
central galaxies.

We conclude that, since different aspects of halo occupation
statistics impact the bias functions $b_\rmg(r_\rmp)$ and $\calR_{\rm
  gm}(r_\rmp)$ in different ways, there is great promise to unveil the
nature of galaxy bias from measurements of $b_\rmg(r_\rmp)$ and
$\calR_{\rm gm}(r_\rmp)$, (or from related quantities, such as the
aperture-based bias parameters introduced by
\citet{1998ApJ...498...43S} and \citet{1998A&A...334....1V}). 
Motivated by 
existing and forthcoming 
imaging and spectroscopic galaxy surveys,
we advocate using a combination of clustering and
galaxy-galaxy lensing to determine the ratio of $\Gamma_{\rm gm} =
b_\rmg/\calR_{\rm gm}$, as a function of the spatial scale $r_{\rm p}$
for a number of different luminosity (and/or stellar mass) bins. 
Inspired by the preliminary work of
\citealt{2004AJ....127.2544S}, we intend to perform such an analysis
in the near future, using data from the SDSS.

\section*{Acknowledgments}
MC has been supported at HU by a Minerva fellowship (Max-Planck
Gesellschaft).  OL acknowledges support of a Royal Society Wolfson
Research Merit Award and a Leverhulme Trust Senior Research
Fellowship. FvdB acknowledges support from the Lady Davis Foundation
for a Visiting Professorship at Hebrew University.

%
\bibliography{paper}
%

\label{lastpage}
\end{document}

%% file: macros.tex
\def\beq{\begin{equation}}
\def\eeq{\end{equation}}
\def\barray{\begin{eqnarray}}
\def\earray{\end{eqnarray}}
\def\beqarray{\begin{eqnarray}}
\def\eeqarray{\end{eqnarray}}

\def\rmc{{\rm c}}
\def\rmd{{\rm d}}

\def\rmg{{\rm g}}
\def\rmh{{\rm h}}

\def\rmm{{\rm m}}

\def\rmp{{\rm p}}

\def\rms{{\rm s}}
\def\rmt{{\rm t}}

\def\rmx{{\rm x}}
\def\rmy{{\rm y}}

\def\rmG{{\rm G}}

\def\calH{{\cal H}}
\def\calI{{\cal I}}

\def\calM{{\cal M}}

\def\calP{{\cal P}}

\def\calR{{\cal R}}

\def\bx{{\bf x}}

\newcommand{\etal}{{et al.~}}

\newcommand{\Mpc}{\>{\rm Mpc}}

\def\gtsima{$\; \buildrel > \over \sim \;$}
\def\ltsima{$\; \buildrel < \over \sim \;$}
\def\prosima{$\; \buildrel \propto \over \sim \;$}
\def\gsim{\lower.7ex\hbox{\gtsima}}
\def\lsim{\lower.7ex\hbox{\ltsima}}
\def\simgt{\lower.7ex\hbox{\gtsima}}
\def\simlt{\lower.7ex\hbox{\ltsima}}
\def\simpr{\lower.7ex\hbox{\prosima}}
\def\la{\lsim}
\def\ga{\gsim}
\def\lta{\la}
\def\gta{\ga}

\newdimen\hssize
\newdimen\hdsize
